\documentclass[aps,pra,reprint,nofootinbib,longbibliography,superscriptaddress,noeprint]{revtex4-2}

\usepackage[T1]{fontenc}
\usepackage{amsmath,amssymb,amsthm}
\IfFormatAtLeastTF{2026-06-01}{}{\usepackage{thmtools}}
\usepackage{graphicx}

\allowdisplaybreaks

\usepackage{mathtools}
\usepackage{physics}
\usepackage{braket}
\usepackage{bm}
\usepackage{microtype}
\usepackage{csquotes}
\usepackage{xcolor}
\usepackage{xurl}
\usepackage[
hyperindex,
breaklinks,
colorlinks=true,
citecolor=teal,
urlcolor=blue]{hyperref}
\usepackage[capitalize]{cleveref}

\newtheorem{theorem}{Theorem}
\newtheorem{lemma}[theorem]{Lemma}
\newtheorem{proposition}[theorem]{Proposition}
\newtheorem{corollary}[theorem]{Corollary}
\theoremstyle{definition}
\newtheorem{remark}{Remark}

\newcommand{\Ext}{\operatorname{Ext}}
\newcommand{\avg}[1]{\langle #1\rangle}

\newcommand{\hs}{\mathcal{H}}
\newcommand{\C}{\mathbb{C}}

\renewcommand{\r}{\right}
\let\polishl\l
\DeclareRobustCommand{\l}{\ifmmode\expandafter\left\else\expandafter\polishl\fi}
\newcommand{\ten}{\otimes}

\renewcommand{\L}{\mathcal{L}}
\newcommand{\Q}{\mathcal{Q}}

\renewcommand{\a}{\alpha}
\renewcommand{\b}{\beta}
\renewcommand{\vb}{\vec{\beta}}
\newcommand{\vP}{\vec{P}}
\newcommand{\p}{\partial}

\newcommand{\vL}{\vec{L}}
\newcommand{\sgn}{\text{sgn}}

\begin{document}

\title{All Bell inequalities with a quantum advantage in the minimal scenario are self-testing}

\author{Mu-En Liu}
\thanks{These authors contributed equally to this work.}
\affiliation{Universit\'e Paris-Saclay, CEA, CNRS, Institut de physique th\'eorique, 91191 Gif-sur-Yvette, France}
\author{Andrea Zingarofalo}
\thanks{These authors contributed equally to this work.}
\affiliation{Universit\'e Paris-Saclay, CEA, CNRS, Institut de physique th\'eorique, 91191 Gif-sur-Yvette, France}
\affiliation{Laboratoire d’Information Quantique, Universit\'e libre de Bruxelles, 1050 Brussels, Belgium}
\author{Victor Barizien}
\affiliation{Department of Applied Physics, University of Geneva, 1211 Geneva, Switzerland}
\author{Jean-Daniel Bancal}
\affiliation{Universit\'e Paris-Saclay, CEA, CNRS, Institut de physique th\'eorique, 91191 Gif-sur-Yvette, France}

\begin{abstract}
In the minimal Bell scenario, all extremal quantum correlations are known, but the Tsirelson inequalities that delimit the quantum set remain poorly understood. We study which correlations can maximize a given Tsirelson inequality, and conversely, which inequalities are maximized by a given correlation. We show that every Tsirelson inequality with quantum advantage is maximized by a single, extremal correlation, and thus self-tests the corresponding realization.
This characterizes all self-tests based on linear Bell inequalities in this scenario.
We also show that at most a two-dimensional family of inequalities can be maximized by a nonlocal extremal correlation. We identify these as polygons for a family of correlations realized by qubit states of arbitrary entanglement. This allows us to deduce that the quantum set is not self-dual.
\end{abstract}

\maketitle

\textit{Introduction.}---Quantum physics predicts the existence of statistics in Bell-type experiments that no local model can reproduce~\cite{Brunner14}. The statistics that can be observed upon measuring a quantum system form the set of quantum correlations---or simply the quantum set $\mathcal Q$. By identifying which statistics are experimentally realizable, this set plays a central role in quantum information science, with applications ranging from foundational questions~\cite{PopescuRohrlich94,Pawlowski09,sun19,pompili26,ramanathan26} to device-independent protocols~\cite{Acin07,Pironio10,Nadlinger22,Zhang22}. Every quantum realization, consisting of a state and local measurements, generates statistics in $\Q$ according to Born's rule. Given some statistics, it is, however, generally a difficult question to determine whether they belong to $\mathcal Q$, or which realization may generate them.

The study of the quantum set was initiated by Tsirelson in the 1980s~\cite{Cirelson80}, and results for the case of statistics with vanishing marginals by Landau and Masanes appeared soon after~\cite{Landau88,Masanes03}. A major step forward came later with the hierarchy of semidefinite programs of Navascu\'es, Pironio and Ac\'in, which approximates the quantum set from the outside~\cite{NPA07,NPA08}. However, at any finite level, this hierarchy does not identify the quantum set exactly, even in the minimal scenario, in which two parties each choose between two binary measurements~\cite{chaturvedi26,pakhunov26}.

Nevertheless, several results have shed light on the geometry of the quantum set. Its boundary was, for instance, shown to contain flat regions that include nonlocal points~\cite{Goh18_Geometry,Rai19,Chen23}, pointy nonlocal extremal points~\cite{customSOS24,Barizien24-extremal} and flat boundaries around local deterministic points~\cite{Cerf2026}. The first complete description of the quantum set in terms of extremal points was recently obtained for the minimal scenario~\cite{Barizien25_CHSH_selftests}. This result also showed that these extremal points self-test, i.e.~identify, the underlying quantum realization~\cite{Supic20-selftesting}.

A description of $\Q$ in terms of its extremal points allows directly computing the maximal quantum value of any Bell expression. However, deciding whether given statistics belong to $\Q$ can be done more directly with an equivalent, dual description in terms of the linear inequalities that delimit its boundary, known as Tsirelson inequalities~\cite{Cirelson80}, the quantum analog of Bell inequalities. Yet the study of this dual picture is still limited. One notable known result is the self-duality property of the quantum set in the minimal scenario with vanishing marginals~\cite{Fritz12}. In the minimal scenario, all the Tsirelson inequalities that are maximized by the Tsirelson point $\vec{P}_T$, the correlation that maximally violates the CHSH inequality~\cite{Clauser69}, were also identified in Ref.~\cite{Barizien24-extremal}. An interesting property of this family of inequalities is that no other nonlocal quantum point can achieve their maximum, i.e.~they all self-test the same realization. The fact that only a single quantum point maximizes a given Tsirelson inequality was also reported in earlier numerical studies~\cite{Goh18_Geometry}.

In this Letter, we ask whether Tsirelson inequalities in the minimal Bell scenario can be maximized by several quantum correlations. For every Tsirelson inequality with a quantum advantage, we show that this is not possible: they can only be maximized by a single nonlocal extremal point. As a consequence, every such inequality self-tests a unique quantum realization. In turn, this allows us to characterize all the self-tests based on Bell inequalities in the minimal scenario. Additionally, we derive further geometrical properties of the quantum set and its dual by studying the orthogonal faces of extremal correlations, the sets of inequalities they maximize. Besides general results, we describe these objects for a specific family of quantum correlations, obtained by measuring partially entangled states. This allows us to show that the quantum set is not self-dual, answering negatively a question raised in~\cite{Fritz12}.

\textit{Preliminaries.}---
We consider the minimal Bell scenario, in which Alice and Bob each choose between two measurements, labeled by $x,y\in\{0,1\}$, with outcomes $\pm1$. 
A correlation is represented by the eight-dimensional vector
$\vP=(\braket{A_x}, \braket{B_y}, \braket{A_xB_y})$.
We denote the quantum set by $\Q$ and the local polytope, the convex hull of the 16 local deterministic correlations, by $\L$. 
One has the evident inclusion that $\mathcal{L} \subset \mathcal{Q}$, and a quantum correlation is said to be nonlocal if it lies outside $\L$.
A Bell functional $\vb \in \mathbb{R}^8$ takes the form
\begin{equation}
\vb\cdot\vP
=\sum_x h_x\langle A_x\rangle
+\sum_y k_y\langle B_y\rangle
+\sum_{x,y}c_{xy}\langle A_xB_y\rangle.
\end{equation}
Its local and quantum bounds are
$\omega_\ell(\vb) \coloneqq \max_{\vP\in\mathcal L}\vb\cdot\vP$ and
$\omega_q(\vb) \coloneqq \max_{\vP\in\Q}\vb\cdot\vP$, respectively.
We say that the correlation $\vP \in \Q$ maximizes $\vb$ if it reaches $\omega_q(\vb)$ and we call $\vb\cdot\vP\leq\omega_q(\vb)$ a \emph{Tsirelson inequality}. 
It exhibits \emph{quantum advantage} when $\omega_q(\vb)>\omega_\ell(\vb)$. In everything that follows, we normalize Bell functionals to have $\omega_q(\vb)=1$.

\begin{figure}[t]
  \centering
  \includegraphics[width=\columnwidth]{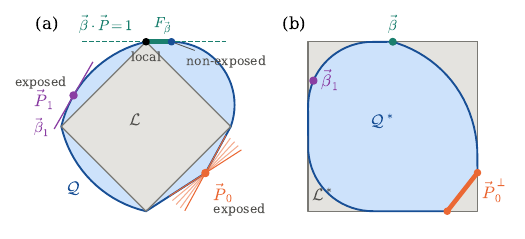}
  \caption{
    Schematic two-dimensional illustration of the notions used in the text. 
    (a) The quantum set $\Q$ (blue) contains the local set $\mathcal L$ (grey). 
    The Tsirelson inequality $\vb$ without quantum advantage is maximized by all correlations on the one-dimensional flat boundary $F_{\vb}$.
    $\vP_0$ is exposed and maximizes a whole family of Tsirelson inequalities (orange), the two extremal ones being also maximized by local points.
    The smooth point $\vP_1$ is also exposed, but it maximizes a single Tsirelson inequality, its tangent $\vb_1$.
    (b) The exact dual of panel~(a), with $\Q^* \subsetneq \L^*$. The inequality $\vb$ is an exposed point of $\mathcal Q^*$ on the boundary of $\mathcal L^*$.
    The inequalities maximized by $\vP_0$ form its orthogonal face $\vP_0^\perp$ (orange), a one-dimensional face of $\Q^*$. The orthogonal face of $\vP_1$ is a single point, i.e.~$\vP_1^\perp=\{\vb_1\}$, which lies in the interior of $\L^*$ since it has quantum advantage.}
  \label{fig:schematic}
\end{figure}

We first start by introducing geometric notions on convex sets. Complete overviews on convex geometry can be found in e.g.~\cite{Rockafellar_ConvexAnalysis}, and all notions discussed here are illustrated in Fig.~\ref{fig:schematic}. Let us first define the dual of the quantum set
\begin{equation}
\Q^* \coloneqq \{\vb\in\mathbb R^8:\vb\cdot\vP\leq1
\text{ for all }\vP\in\Q\}, 
\end{equation}
which is the set of all Tsirelson inequalities. Note that due to duality, one has $\mathcal{Q}^* \subset \mathcal{L}^*$. A Bell functional has quantum advantage if it is in the interior of $\mathcal{L}^*$. The Tsirelson inequalities maximized by a given correlation $\vP$ form its \emph{orthogonal face}
\begin{equation}
\label{eq:orth-face}
\vP^\perp \coloneqq \{\vb\in\Q^*:\vb\cdot\vP=1\},
\end{equation}
which is a face of $\mathcal{Q}^*$ and whose affine dimension $\dim(\vP^\perp)$ is called the \emph{orthogonal dimension}. Conversely, the set of correlations that maximize a given Bell functional $\vb$, given by
\begin{equation}\label{eq:face}
  F_{\vb}\coloneqq\{\vP\in\Q:\vb\cdot\vP=1\},
\end{equation}
is a face of $\Q$. We call $F_{\vb}$ a \emph{flat boundary} of $\Q$ when it contains more than one point, i.e.~when $\dim(F_{\vb}) \geq 1$. A point $\vP$ of $\Q$ is \emph{extremal} if it admits no nontrivial convex decomposition into other quantum correlations, and \emph{exposed} if it uniquely maximizes some functional $\vb \in \Q^*$, i.e.~if $F_{\vb} = \{\vP\}$. Similarly, a Tsirelson inequality is extremal if it is not a nontrivial convex mixture of other Tsirelson inequalities. The extremal Tsirelson inequalities are the tight inequalities of $\Q$, as the CHSH inequalities and the positivity constraints are for the local polytope.

\textit{Tsirelson inequalities with quantum advantage admit a unique maximizer.}---
A priori, it is conceivable that a given Tsirelson inequality can be maximized by several different extremal points. For instance, a Tsirelson inequality which does not have quantum advantage might be maximized by a nonlocal extremal point and one or more local deterministic points. The question is thus whether a Tsirelson inequality with quantum advantage can be maximized by several nonlocal extremal points.

Strikingly, it was found in Ref.~\cite{Barizien24-extremal} that all Tsirelson inequalities with quantum advantage that are maximized by the Tsirelson point $\vP_T$ are only maximized by $\vP_T$ itself.
Hence, all of them admit a single nonlocal maximizer.
It is then natural to ask whether this is due to the fact that $\vP_T$ is generated by a maximally entangled state.

Surprisingly, we subsequently found that all Tsirelson inequalities with quantum advantage maximized by a family of extremal correlations that do not arise from maximally entangled states (see Appendix~\ref{app:fam-family}) are again only maximized by the point itself.
We then conclude that maximally entangled states are not the determining factor for the phenomenon of uniqueness of the maximizer, which might be more ubiquitous in the minimal scenario.
This intuition is formalized in the following lemma.
\begin{lemma}
\label{thm:main}
  In the minimal scenario, any Tsirelson inequality that admits a quantum advantage is maximized by exactly one nonlocal extremal correlation in the quantum set $\Q$. 
  Moreover, this correlation is exposed.
\end{lemma}

The proof of~\cref{thm:main}, detailed in Appendix~\ref{app:uniqueness}, relies on two key results.
First, we show that a Tsirelson inequality, even with marginal terms, exhibits quantum advantage only if the product of the correlator coefficients satisfies $c_{00}c_{01}c_{10}c_{11}<0$. Note that a sufficient and necessary condition for a Tsirelson inequality that only involves correlators to exhibit quantum advantage has been derived in~\cite{Ramanathan17}. This requirement on the correlator coefficients is enough to fix the qubit measurements that maximize the inequality up to local unitaries.
Secondly, we prove that the state reaching the quantum bound of any Bell operator with quantum advantage is unique. An important feature in the minimal scenario is that all extremal correlations can be realized by pure two-qubit states and projective measurements~\cite{Masanes03,Barizien25_CHSH_selftests}. This ensures that the Bell operator and its eigenvectors are in a two-qubit space. Restricting the dimension of the realization guarantees that the eigenspace of the largest eigenvalue of any Bell operator associated with a Tsirelson inequality with quantum advantage cannot be two-dimensional, as any two-dimensional subspace in $\C^2 \ten \C^2$ must contain a product state~\cite{Parthasarathy2004}, and thus the quantum bound of this Bell operator would equal the local bound.
Altogether, the two facts then imply that the nonlocal extremal maximizer is unique.
Furthermore, this extremal maximizer is the only maximizer, since no local point reaches the quantum bound of an inequality with quantum advantage. Hence, it is exposed.

\cref{thm:main} has direct consequences for the geometry of the quantum set. Indeed, it implies that any flat boundary $F_{\vb}$ of $\mathcal{Q}$ (which has at least two extremal points) must contain a local deterministic point.
\begin{corollary}
    All flat boundaries of the quantum set in the minimal scenario intersect with the local polytope.
\end{corollary}

\textit{All Tsirelson inequalities with quantum advantage are useful for self-testing.}---
By~\cref{thm:main}, a Tsirelson inequality with quantum advantage reaches its maximum at a single nonlocal extremal quantum correlation. In the minimal scenario, every such correlation self-tests its quantum realization~\cite{Barizien25_CHSH_selftests}, i.e.~all realizations that produce it are equivalent up to local isometries. Thus, observing the maximal value of the inequality is enough to self-test this realization.
\begin{theorem}
\label{thm:selftest}
    All Tsirelson inequalities with quantum advantage in the minimal scenario self-test their unique maximizer.
\end{theorem}
Tsirelson inequalities without quantum advantage are maximized by a local deterministic point, so they cannot self-test a nonlocal realization. 
This result thus provides a complete characterization of all Tsirelson inequalities whose maximal value self-tests a quantum realization.

\textit{The geometry of the orthogonal faces.}---
Studying the orthogonal faces of the extremal points of the quantum set is of key interest to describe its geometry and its dual. Indeed, every point on the boundary of $\mathcal{Q}^*$ (and thus every extremal Tsirelson inequality) must be maximized by at least one extremal point of $\mathcal{Q}$. In Ref.~\cite{Barizien25_CHSH_selftests}, it was established that extremal quantum correlations fall into one of three categories: (i) local deterministic strategies; (ii) nonlocal non-exposed points that lie on the boundary of the no-signaling set; (iii) nonlocal points not touching the no-signaling border. Points in (i) saturate twelve positivity constraints and have maximal seven-dimensional orthogonal faces. Following Ref.~\cite{Barizien24-extremal}, one can upper bound the dimension of the orthogonal face of a point in class (ii) or (iii) by considering small variations of state and measurement parameters of the underlying quantum realization, and asking for a non-increasing Bell functional. In Appendix~\ref{app:dimensionbound}, we conduct such analysis for all nonlocal extremal correlations, and show the following.
\begin{lemma}
\label{thm:dual}
  The orthogonal face of every nonlocal extremal correlation in $\Q$ has dimension at most two.
\end{lemma}
\noindent In other words, the set of Tsirelson inequalities $\vb$ maximized by a nonlocal extremal correlation $\vP$ can be parametrized by at most two parameters, allowing us to plot the orthogonal face of $\vP$ in a two-dimensional plane. Studying such planes is of practical interest, as it provides parametrized families of Tsirelson inequalities for self-testing a given realization, with potentially different noise robustness. Note that it was also proven in Ref.~\cite{Barizien25_CHSH_selftests} that all inequalities maximized by points in (ii) have no quantum advantage. As such, for points in (i) and (ii), it holds that $\vP^\perp \subset \partial \mathcal{L}^*$. This is not the case anymore for points in class (iii) which can be exposed, and thus have Tsirelson inequalities with quantum advantage in their orthogonal face.

\begin{figure}[t]
  \centering
  \includegraphics[width=\columnwidth]{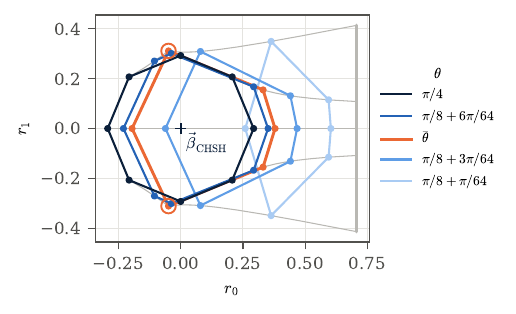}
  \caption{The orthogonal faces $\vec P_\theta^\perp$ for the extremal points $\vec P_\theta$ can be parametrized by two coordinates $r_0$ and $r_1$ for several values of $\theta\in(\pi/8,\pi/4]$, see Appendix~\ref{app:fam-family}. 
  For all $\pi/8 < \theta \leq \pi/4$, the orthogonal face is a polytope, and the inequalities in its interior admit a quantum advantage, while those on the boundary are also maximized by local deterministic points.
  The set $\vP_\theta^\perp$ is a hexagon below a critical value $\bar\theta$ and an octagon above it.
  When $\theta=\pi/4$, the set is a regular octagon centered at the CHSH inequality $\vec\beta_{\rm CHSH}$. 
  When $\theta\to\pi/8$, it collapses onto the grey segment.}
  \label{fig:faces}
\end{figure}

We conducted an analysis of the orthogonal faces for a specific family of quantum correlations $\vec P_\theta$ in class (iii), obtained from partially entangled two-qubit states $\ket{\phi_\theta} = \cos(\theta)\ket{00} + \sin(\theta)\ket{11}$ measured with $ A_x = (\sigma_z + (-1)^x\sigma_x)/\sqrt{2}$ for Alice and $B_0 = \sigma_z$, $B_1=\sigma_x$ for Bob, where $\sigma_x, \sigma_z$ denote the Pauli $x$ and $z$ operators. These correlations are extremal for all values of $\theta\in[\pi/8,\pi/4]$ and exposed whenever $\theta\neq\pi/8$. 
For $\theta=\pi/4$, one recovers the Tsirelson point $\vP_T$, whose orthogonal face was described in Ref.~\cite{Barizien24-extremal}. In Appendix~\ref{app:fam-family}, we give a full description of the orthogonal face of these points for the full entanglement range, using reduction to a two-dimensional plane by local variations, sum-of-squares certificates, and local bounds to respectively inner and outer approximate the exact region of the plane where the quantum bound is exactly $1$. The orthogonal faces are plotted in Fig.~\ref{fig:faces} for different values of $\theta$. Strikingly, we obtain that for all $\theta\in(\pi/8,\pi/4]$, the orthogonal face is two-dimensional and forms a polytope whose extremal points have no quantum advantage. In other words, for the points studied here, it still holds that $\text{Ext}(\vec P_\theta^\perp) \subset \partial \mathcal{L}^*$. Furthermore, we observe that for values of $\theta$ larger than some threshold value $\bar\theta\approx0.645$, the orthogonal face is an octagon, regular only at $\theta=\pi/4$, while below the critical value, $\pi/8 < \theta \leq \bar\theta$, the orthogonal face is a hexagon. At the limit point $\theta=\pi/8$, the point actually belongs to class (ii), i.e.~is non-exposed, and the orthogonal face is only one-dimensional, with its two extremal points being positivity constraints.

For $\theta\in(\pi/8,\pi/4]$, only the eight local deterministic points saturating the CHSH inequality are relevant to the identification of the boundary of the orthogonal face. It turns out that for $\theta\neq\bar\theta$, each vertex of $\vP_\theta^\perp$ is a Tsirelson inequality maximized by $\vP_\theta$ and two deterministic points. These vertices are extremal in $\Q^*$ and have orthogonal dimension $\dim F_{\vb}\geq 2$. However, at $\bar\theta$, two pairs of adjacent vertices merge, and the merged vertices $\vb^\star_1$, $\vb^\star_2$ are maximized by $\vP_{\bar\theta}$ and three deterministic points. These four points are proven to be affinely independent, so these two inequalities are extremal points of $\mathcal{Q}^*$ with orthogonal dimension $\dim F_{\vb^\star_i}\geq 3$. This observation leads to the following result.
\begin{theorem}[No self-duality]
\label{thm:not-self-dual}
  No linear map $T$ satisfies $T(\Q)=\Q^*$, so $\Q$ is not self-dual.
\end{theorem} 
The proof, presented in Appendix~\ref{app:not-self-dual}, relies on the fact that any such linear map $T$ with $T(\Q)=\Q^*$ is invertible, maps extremal points to extremal points, and preserves orthogonal dimensions. As such, we prove that it would need to map $F_{\vec\beta^\star_i}$ onto the orthogonal face of a nonlocal extremal point of $\Q$ (a counting argument on the positivity constraints rules out the deterministic points), whose dimension is at most two according to~\cref{thm:dual}. Hence no such $T$ exists.

Note that in the subspace of vanishing marginals ($\braket{A_x}= \braket{B_y}=0$), the quantum set is self-dual~\cite{Fritz12,Le23}. While it was already known that analogous self-duality fails for three parties~\cite{Fritz12}, \Cref{thm:not-self-dual} shows that it also fails when only two parties are involved once the marginals are included. This shows that even in the minimal Bell scenario, the 
geometries of the quantum set of correlations $\mathcal{Q}$ and of its dual $\mathcal{Q}^*$, the set of Tsirelson inequalities, differ. 

\textit{Conclusion.}---
In this Letter, we show that every Tsirelson inequality in the minimal scenario that admits quantum advantage is maximized by a single nonlocal and exposed quantum correlation.
In turn, this implies that every such inequality is useful for self-testing the realization of its maximizer, and that all flat boundaries of the quantum set intersect with the local polytope. 
Furthermore, we find that the orthogonal face of every nonlocal extremal correlation has dimension at most two, so the Tsirelson inequalities maximized by a given nonlocal extremal correlation form a family with at most two parameters. We further described such orthogonal faces for a family of extremal and exposed quantum correlations obtained by measuring a partially entangled state. We showed that for this family, the orthogonal faces are two-dimensional polytopes whose vertices are maximized by two or three local deterministic correlations. This provides both a practical scope for certifying these realizations with noise-tailored inequalities, as well as theoretical implications for the study of the quantum set and its dual. For instance, finding extremal Tsirelson inequalities maximized by four affinely independent extremal correlations allowed us to show that the quantum set is not self-dual.

While our Letter provides great insight for describing the dual of the quantum set in the minimal scenario, it also calls for description beyond. In general Bell scenarios, the eigenspace of the largest eigenvalue of a Bell operator with quantum advantage still contains no product state. In $\mathbb C^d\otimes\mathbb C^{d'}$, however, such completely entangled subspaces can have dimension up to $(d-1)(d'-1)$~\cite{Parthasarathy2004}, so our argument does not easily extend beyond two qubits.
In fact, the uniqueness of the nonlocal maximizer is known not to extend to general Bell scenarios: Bell inequalities with quantum advantage can admit multiple maximizing nonlocal correlations~\cite{Hsu26}.
Thus, not all Tsirelson inequalities with quantum advantage can be used to self-test correlations. Our work still raises the question of whether it would be possible to find necessary conditions for uniqueness.

\textit{Acknowledgements.}---
MEL thanks Kai-Siang Chen and Yeong-Cherng Liang for helpful discussions.
We acknowledge funding from the Taiwan Paris-Saclay Doctoral Scholarships Program, and support from CEA. This work was also supported by QuantERA ToDiQT, by the French National Research Agency (ANR) under grant number ANR-26-QUA2-0011-05, and co-funded by the European Commission. AZ is a Research Fellow of the Fonds de la Recherche Scientifique -- FNRS, and acknowledges funding from the FWO and F.R.S.-FNRS under the Excellence of Science (EOS) programme project 40007526. VB acknowledges funding from the Swiss National Science Foundation (project 219366).

\textit{Code availability.}---The Python scripts that produce the numbers and figures of Appendix~\ref{app:fam-family}, and the Mathematica code for the exact proofs and the sum-of-squares certificates, are available at \url{https://doi.org/10.5281/zenodo.23956844}.

\textit{AI Statement.}---OpenAI’s ChatGPT (GPT-6 Astra) and Anthropic's Claude Opus 5.5 were used to assist in deriving the proof of \Cref{thm:main}, developing the sum-of-squares certificates presented in Appendix~\ref{app:fam-sos}, and preparing the manuscript. All proofs, claims, and conclusions have been reviewed and verified by the authors, who take full responsibility for this work.

\textit{Note added.}---While completing this project, we became aware of a related work by L.-L. Sun, J. Wang and S. Yu~\cite{sun26}.

\bibliography{Ref.bib, Ref_family.bib}

\clearpage
\onecolumngrid
\makeatletter
\onecolumn@grid@setup
\let\set@footnotewidth\set@footnotewidth@one
\let\compose@footnotes\compose@footnotes@one
\makeatother

\appendix
\renewcommand{\theHequation}{\thesection.\arabic{equation}}

\section{Proof of~\cref{thm:main}}\label{app:uniqueness}

\subsection{Preliminaries: considerations on the quantum set} \label{app:preliminaries}

We express the correlation $\vP \in \mathbb{R}^8$ in the minimal scenario as a vector with components in the following order:
\begin{eqnarray}
\label{eq:vec-order}
    \vP \coloneqq
    \begin{pmatrix}
        \braket{A_0}, \braket{A_1}, \braket{B_0}, \braket{B_1}, \braket{A_0B_0}, 
        \braket{A_0B_1},
        \braket{A_1B_0},
        \braket{A_1B_1}
    \end{pmatrix}.
\end{eqnarray}
It is known~\citep[Proposition A.1]{Barizien25_CHSH_selftests} that the extremal correlation $\vP$ can always be parametrized with five parameters, $(\theta, a_0, a_1, b_0, b_1)$ if we take the following strategy:
\begin{equation} \label{eq:realizqubit}
\begin{aligned}
    \ket{\phi_\theta} &\coloneqq \cos{\theta}\ket{00} + \sin{\theta}\ket{11}, \\
    A_x &\coloneqq \cos(a_x)\sigma_z + \sin(a_x)\sigma_x, \\
    B_y &\coloneqq \cos(b_y)\sigma_z + \sin(b_y)\sigma_x.
\end{aligned}
\end{equation}
Specifically, for this realization, one has
\begin{equation}
\label{eq:qubit-ext-corr}
    \braket{A_x} = \cos(a_x) \cos(2\theta), \quad
    \braket{B_y} = \cos(b_y) \cos(2\theta), \quad
    \braket{A_xB_y} = \sin(a_x)\sin(b_y) \sin(2\theta) + \cos(a_x)\cos(b_y).
\end{equation}
For convenience, we will sometimes express the vectors throughout the paper in the following manner:
\begin{equation}
    \vP 
    =  
    \l(\begin{array}{c|c|c}
        & \braket{B_0} 
        & \braket{B_1} \\ \hline
        \braket{A_0} 
        & \braket{A_0B_0} 
        & \braket{A_0B_1} \\ \hline
        \braket{A_1}
        & \braket{A_1B_0} 
        & \braket{A_1B_1}
    \end{array}\r)\\
    =  
    \l(\begin{array}{c|c|c}
        & c_{b_0} c_{2\theta} 
        & c_{b_1} c_{2\theta} \\ \hline
        c_{a_0} c_{2\theta} 
        & s_{a_0}s_{b_0}s_{2\theta} + c_{a_0}c_{b_0} 
        & s_{a_0}s_{b_1}s_{2\theta} + c_{a_0}c_{b_1}  \\ \hline
        c_{a_1} c_{2\theta}
        & s_{a_1}s_{b_0}s_{2\theta} + c_{a_1}c_{b_0} 
        & s_{a_1}s_{b_1}s_{2\theta} + c_{a_1}c_{b_1}
    \end{array}\r),
\end{equation}
where standard abbreviations for trigonometric functions are used, e.g.~$c_{\theta} \coloneqq \cos(\theta)$ and $s_{\theta} \coloneqq \sin(\theta)$.

Two correlations $\vP(\theta,a_0,a_1,b_0,b_1)$ and $\vP'(\theta',a_0',a_1',b_0',b_1')$ are {\em operationally equivalent} if they can be transformed into one another by the following relabelings:
\begin{enumerate}
    \item Party relabeling: Swap $a_0$ with $b_0$ and swap $a_1$ with $b_1$.
    \item Input relabeling: Swap $a_0$ with $a_1$ or swap $b_0$ with $b_1$.
    \item Output relabeling: Transform $A_x (B_y)$ to $-A_x (-B_y)$ for some $x(y)$.
\end{enumerate}
Here, we show that any extremal correlation can be transformed into an operationally equivalent correlation with parameters in a specific range.
\begin{lemma}
    Any extremal point in the minimal scenario can be relabeled to have the parameters in the following range.
    \begin{equation}
    \label{eq:range}
        0 \leq \theta \leq \pi/4,  \quad 
        a_0 \leq b_0, \quad
        a_0 \leq a_1, \quad 
        b_0 \leq b_1, \quad 
        a_x,b_y \in [0,\pi).
    \end{equation}
\end{lemma}
\begin{proof}

Suppose that $a_x$ and $b_y$ are in the range of $[\pi,2\pi)$. One can augment a minus sign in front of $A_x$ and $B_y$, which corresponds to sending $a_x$ to $a_x - \pi$ and $b_y$ to $b_y - \pi$. By doing so, we can restrict $a_x$ and $b_y$ to be in $[0,\pi)$ by relabeling the outputs. We can also relabel the indices of the parties to ensure the relation $a_0 \leq b_0$. Also, input relabeling allows us to ensure $a_0 \leq a_1$ and $b_0 \leq b_1$.

We now show that the Schmidt angle $\theta$ can be restricted to lie within $[0, \pi/4]$ by relabeling the parties and the measurement outcomes.
Assume that $\theta$ is within $[\pi/2, \pi]$.
Let us rewrite the state $\ket{\phi(\theta)}$ so that its Schmidt angle is within $[0, \pi/2]$.
\begin{equation}
    \begin{aligned}
        \ket{\phi(\theta)} 
        &= \cos(\theta) \ket{00} + \sin(\theta) \ket{11} \\ 
        &= -\Big( 
            \cos(\pi - \theta) \ket{00} 
            - \sin(\pi - \theta) \ket{11}
            \Big)\\
        &= - (\openone \ten \sigma_z) \ket{\phi(\pi - \theta)}.
    \end{aligned}
\end{equation}
After this, the state $\ket{\phi(\theta')} = \ket{\phi(\pi - \theta)}$ would have an angle $\theta'$ lying in $[0, \pi/2]$.
Now, assume that $\theta \in [\pi/4, \pi/2]$.
The following operation brings the state's Schmidt angle to the desired range.
\begin{equation}
    \begin{aligned}
        \ket{\phi(\theta)} 
        &= \cos(\theta) \ket{00} + \sin(\theta) \ket{11} \\ 
        &= 
        \sin(\pi/2 - \theta) \ket{00} +
        \cos(\pi/2 - \theta) \ket{11}\\
        &= (\sigma_x \ten \sigma_x) \ket{\phi(\pi/2- \theta)}.
    \end{aligned}
\end{equation}
Note that $\pi/2 - \theta$ lies in $[0, \pi/4]$ now.
The operations on the state, $\openone \ten \sigma_z, \sigma_x \ten \sigma_x$, are nothing but modifications to the measurement settings; 
$\openone \ten \sigma_z$ takes $b_y$ to $-b_y$, and 
$\sigma_x \ten \sigma_x$ takes $a_x(b_y)$ to $\pi - a_x(b_y)$.
Thus, after the operation, one can relabel the inputs, outputs and party indices again to retain the relation $a_0 \le a_1, b_0 \le b_1, a_0 \le b_0$.
\end{proof}

For convenience, let us introduce the half-angle variables:
\begin{equation}
\label{eq:half-angles}
\a_x \coloneqq \tan(a_x/2), \quad 
\b_y \coloneqq \tan(b_y/2), \quad
t \coloneqq \tan(\theta),
\end{equation}
Within the range of~\cref{eq:range}, they can be further restricted to:
\begin{equation}
\label{eq:half-angles-range}
0 \leq t \leq 1, \quad 
0 \leq \a_x, \b_y, \quad
\forall x,y.
\end{equation}
We now recall the characterization of extremal correlations in the minimal scenario stated in~\citep[Theorem 1]{Barizien25_CHSH_selftests}.
\begin{proposition}
\label{prop:FA}
A nonlocal correlation $\vP$ is extremal if and only if it admits an operationally equivalent point $\vP'$ of the form~\cref{eq:qubit-ext-corr}, with parameters in the range~\cref{eq:range}, satisfying all the conditions below:
\begin{equation}
\label{eq:extremal-corr}
\a_0 \leq \b_0 t, \quad
\b_0 \leq \a_1 t, \quad
\a_1 \leq \b_1 t.
\end{equation}
\end{proposition}
We say that an extremal correlation $\vP$ satisfies {\em strict full alternation} if all the inequalities in~\cref{eq:extremal-corr} are strict.
In the proof of~\citep[Lemma 2]{Barizien25_CHSH_selftests}, violation of strict full alternation is shown to be a sufficient condition for the nonlocal extremal correlation to be non-exposed.
\begin{proposition}
\label{prop:not-strict-FA}
    Consider a nonlocal extremal correlation $\vP$ in the minimal scenario.
    If it saturates one of the inequalities in~\cref{eq:extremal-corr}, then there exists a local correlation $\vL$ that maximizes all the Tsirelson inequalities $\vb$ maximized by $\vP$. As a result, $\vP$ is non-exposed.
\end{proposition}
Although known examples of nonlocal extremal correlations that are non-exposed violate strict full alternation (see, for example,~\citep[Appendix H]{Chen23} and the Hardy correlation~\cite{Hardy93}), it is not clear if this is a necessary condition for non-exposedness. However, only points with strict full alternation can maximize a Tsirelson inequality with quantum advantage.

\begin{lemma}
\label{lem:adv-fa}
  Let $\vb$ be a Tsirelson inequality with quantum advantage. 
  Then, the extremal correlations that maximize $\vb$ must satisfy strict full alternation.
\end{lemma}
\begin{proof}
    Suppose that $\vb$ is maximized by a nonlocal extremal correlation $\vP$ that does not satisfy strict full alternation. \cref{prop:not-strict-FA} states that there exists a local correlation that can also maximize $\vb$, contradicting the assumption that $\vb$ has quantum advantage.
\end{proof}

\subsection{A necessary condition on the correlator coefficients for quantum advantage}

\begin{lemma}
\label{lem:pxy-strict-bound}
    Consider a correlation $\vP(\theta,a_0,a_1,b_0,b_1)$, and define:
    \begin{equation}
    \label{eq:pxy-definition}
        p_{xy} \coloneqq\cos(a_x)\cos(b_y)+\frac{\sin(a_x)\sin(b_y)}{\sin(2\theta)}.
    \end{equation}
    If $\vP$ is nonlocal and satisfies strict full alternation, then $\abs{p_{xy}} < 1$ for all $x,y \in \{0,1\}$.
\end{lemma}
\begin{proof}
We can rewrite~\cref{eq:pxy-definition} into
\begin{equation}
\label{eq:pxy-factorizations}
\begin{aligned}
    p_{xy}+1&=\frac{2(\a_x\b_y+t)(1+\a_x\b_y t)}{t(1+\a_x^2)(1+\b_y^2)},\\
    p_{xy}-1&=\frac{2(\b_y-\a_x t)(\a_x-\b_y t)}{t(1+\a_x^2)(1+\b_y^2)}.
\end{aligned}
\end{equation}
Nonlocality requires $0<t\leq 1, \ \a_x,\b_y\geq 0$. In particular, $t(1+\a_x^2)(1+\b_y^2)>0$, so the denominators are well defined and strictly positive. Also, nonlocality ensures that both factors in the numerator of the first expression are strictly positive, implying that $p_{xy}>-1$ for every $x,y\in\{0,1\}$.

It remains to prove that $p_{xy}<1$. First, consider $(x,y)\in\{(0,0),(0,1),(1,1)\}$. For these three pairs, strict full alternation in~\cref{eq:extremal-corr} implies $\a_x<\b_y t$. For $(x,y)=(0,1)$ this follows from $\a_0<\b_0t<\a_1t^2<\b_1t^3\leq\b_1t$. Since $0<t\leq1$, we also have $\a_x t<\b_y t^2\leq\b_y$, and therefore $\b_y-\a_x t>0$. Thus,$(\b_y-\a_x t)(\a_x-\b_y t)<0$ for these pairs. For the remaining pair $(x,y)=(1,0)$, strict full alternation requires $\b_0-\a_1t<0$. Multiplying it by $t$ and using $t\leq1$ yields $\b_0t<\a_1t^2\leq\a_1$, so $\a_1-\b_0t>0$. Hence, $(\b_0-\a_1t)(\a_1-\b_0t)<0$. We have therefore shown that
\begin{equation}
    (\b_y-\a_x t)(\a_x-\b_y t)<0
\end{equation}
for every $x,y\in\{0,1\}$. The second identity in~\cref{eq:pxy-factorizations} then gives $p_{xy}<1$. Combining the two bounds gives $-1<p_{xy}<1$ for every $x,y\in\{0,1\}$.
\end{proof}

The most general Bell expression $\vb$ in the minimal scenario can be written as
\begin{equation}
\label{eq:bell-functional}
  \vb \cdot \vP =
  \sum_{x=0}^1h_x\avg{A_x} +
  \sum_{y=0}^1k_y\avg{B_y} +
  \sum_{x,y=0}^1c_{xy}\avg{A_xB_y},
\end{equation}
where all coefficients are real. We first establish a necessary condition for a Bell functional to be maximized by a nonlocal extremal correlation that satisfies strict full alternation.

\begin{lemma}\label{lem:sign}
Consider $\vb$ that is maximized by a nonlocal extremal correlation $\vP$.
If $\vP$ satisfies strict full alternation, then the correlator coefficients of $\vb$ defined in~\cref{eq:bell-functional} satisfy
\begin{equation}
\label{eq:negative-product}
  c_{00}c_{01}c_{10}c_{11}<0.
\end{equation}
In particular, none of the four correlator coefficients vanishes.
\end{lemma}

\begin{proof}
Observe that we can eliminate the marginal terms in $\vP$ by adding one of the tangent vectors. Explicitly,
\begin{equation}
\label{eq:marginal-elimination}
  \vP+\frac{\cot2\theta}{2}\,\p_\theta \vP
  =(0,0,0,0,p_{00},p_{01},p_{10},p_{11}),
\end{equation}
where $p_{xy}$ is defined in~\cref{eq:pxy-definition}.
For a nonlocal correlation $\vP$, the expression is well defined as we always have $\theta \neq 0$.
Since small variation of the Schmidt angle $\theta$ should not increase the Bell value, we have at first order $\vb \cdot \p_\theta \vP = 0$, and in particular
\begin{equation}
  \omega_q
  = \vb \cdot \Big(\vP + \dfrac{1}{2} \cot(2\theta)\p_\theta \vP \Big)
  = \sum_{x,y} c_{xy} p_{xy}.
\end{equation}
Suppose that $c_{00}c_{01}c_{10}c_{11} > 0$.
Then, $\sgn(c_{00}) \sgn(c_{11}) = \sgn(c_{01}) \sgn(c_{10})$. Consider the correlation with vanishing marginals and correlators $E_{xy}\coloneqq\sgn(c_{xy})$. Since $E_{00}E_{11}=E_{01}E_{10}$, it is the equal mixture of two opposite deterministic points, so it is local.
Hence,
\begin{equation}
  \sum_{x,y} c_{xy} E_{xy} =
  \sum_{x,y} \abs{c_{xy}} > \sum_{x,y} c_{xy} p_{xy}.
\end{equation}
The last inequality follows from~\cref{lem:pxy-strict-bound}, which establishes that strict full alternation forces $\abs{p_{xy}}<1, \forall x,y$.
This contradicts the assumption that $\vP$ maximizes the Bell functional.
The same argument applies if some, but not all, of the coefficients vanish. If all of them vanish, $\vb$ only has marginal terms. Since $\cos2\theta<1$, $\vb\cdot\vP\leq\cos(2\theta)\big(\sum_x|h_x|+\sum_y|k_y|\big)$ is then strictly below the value $\sum_x|h_x|+\sum_y|k_y|$ of a deterministic point, so $\vP$ cannot maximize $\vb$.
\end{proof}

\begin{corollary}
\label{coro:Q-advantage}
  In the minimal scenario, if $\vb$ has quantum advantage, then the correlator coefficients of $\vb$ satisfy $c_{00}c_{01}c_{10}c_{11}<0$, even when marginal terms are present.
\end{corollary}
\begin{proof}
    If $\vb$ has quantum advantage, then, by~\cref{lem:adv-fa}, it is maximized by a nonlocal extremal correlation that satisfies strict full alternation.
    By \cref{lem:sign}, the correlator coefficients of such $\vb$ satisfy $c_{00}c_{01}c_{10}c_{11}<0$.
\end{proof}

\subsection{Uniqueness of the realization parameters}

The Bell operator that corresponds to the Bell expression~\cref{eq:bell-functional} for a given choice of measurements is
\begin{equation}
  \hat S
  \coloneqq
  \sum_{x=0}^1 h_x A_x
  + \sum_{y=0}^1 k_y B_y
  + \sum_{x,y=0}^1 c_{xy} A_x B_y.
\end{equation}
For every strategy, the Bell operator satisfies $\openone-\hat S\succeq 0$, and a strategy attains $\omega_q$ when its state is an eigenvector of $\hat S$ with eigenvalue $\omega_q$. In the minimal scenario, these strategies are ensured to be realizable by qubit projective and real measurements on a pure two-qubit state (\citep[Proposition A.1]{Barizien25_CHSH_selftests}).
Hence, when evaluating $\openone-\hat S\succeq 0$, it suffices to consider such strategies. Since we only care about the maximal eigenvalue of the Bell operator, we can reduce the number of parameters by rotating the observables with local unitaries to
\begin{equation}
  A_0=Z,
  \qquad
  A_1=uZ+\sqrt{1-u^2}\,X,
  \qquad
  B_0=Z,
  \qquad
  B_1=vZ+\sqrt{1-v^2}\,X,
\end{equation}
where $u,v$ characterize the measurement overlaps of the two observables of each party, as $u\coloneqq\cos(a_0 - a_1)$ and $v\coloneqq\cos(b_0 - b_1)$.
The relevant Bell operator that attains the quantum bound is then a function of the overlaps $u$ and $v$ only: $\hat S = \hat S(u,v)$. Note that the same parametrization technique was also used before, e.g.~to parametrize the Bell operator of the doubly tilted CHSH inequality in~\cite{GigenaEtAl2025}.

\begin{lemma}
\label{lem:overlap-unique}
Consider a Tsirelson inequality $\vb$ with quantum advantage.
Suppose that the nonlocal extremal correlation $\vP_1$ maximizes $\vb$ with measurement overlaps $(u,v)$.
Then, for any other nonlocal extremal point $\vP_2$ with overlaps $(u',v')$ that also maximizes the same Bell functional, it must be the case that $(u',v')=(u,v)$. 
Moreover, the eigenspace of $\hat S(u,v)$ associated with its largest eigenvalue is one-dimensional.
\end{lemma}

\begin{proof}
Define
\begin{equation}
\label{eq:D-direct}
  D(u,v) \coloneqq \det \big(\openone-\hat S(u,v) \big).
\end{equation}
Since $\openone-\hat S(u,v)\succeq0$ for every $(u,v)$, we have $D(u,v)\geq0$ on $[-1,1]^2$. A direct symbolic calculation shows that $D$ is a polynomial of degree at most two in $u$ and at most two in $v$, in which $u^2v^2$ has the coefficient $16c_{00}c_{01}c_{10}c_{11}$. 
Moreover, the roots of $D(u,v) = 0$ are precisely the overlap pairs attaining the quantum bound. We now show that $D(u,v)=0$ has at most one solution. 

Suppose that for some $(u_1,v_1), (u_2,v_2)$, one has $D(u_1,v_1)=D(u_2,v_2)=0$. First, assume that $u_2\neq u_1$ and $v_2\neq v_1$.
We evaluate $D(u,v)$ along the line segment joining $(u_1,v_1)$ and $(u_2,v_2)$ by defining
\begin{equation}
\label{eq:f-line}
    f(s)\coloneqq
    D(u_1+s\Delta u,v_1+s\Delta v),
\end{equation}
where $s\in[0,1]$,
$\Delta u\coloneqq u_2-u_1$, and $\Delta v\coloneqq v_2-v_1$.

Since $f(0)=f(1)=0$, both endpoints are roots of
$f(s)$. 
Nonlocality requires incompatible local observables, which implies that $u_1,u_2,v_1,v_2$ lie strictly inside $(-1,1)$.
Together with $D(u,v)\geq0$ on $[-1,1]^2$, this ensures that $f(s)$ is nonnegative on both sides of each root. 
Thus, unless $f = 0$ for all $s$, both roots have even multiplicity.
A direct calculation shows that $f(s)$ has degree at most four. 
Therefore, $f(s)=\gamma s^2(s-1)^2$, where $\gamma$ is the coefficient of $s^4$, given by
\begin{equation}
    \gamma
    =
    16c_{00}c_{01}c_{10}c_{11}
    (\Delta u)^2(\Delta v)^2.
\end{equation}
By assumption, $\vb$ has quantum advantage, so \cref{coro:Q-advantage} implies that $c_{00}c_{01}c_{10}c_{11}<0$. Since $\Delta u,\Delta v\neq0$, we obtain $\gamma<0$. In particular, $f(s)<0$ for $s\in(0,1)$, contradicting its nonnegativity.

It remains to consider the cases in which exactly one overlap coincides. Without loss of generality, suppose that $v_2=v_1$ but $u_2\neq u_1$, and define
\begin{equation}
\label{eq:second-fs}
    g(s)\coloneqq D(u_1+s\Delta u,v_1).
\end{equation}
If $D(\cdot,v_1)$ is not identically zero, then $g$ is a nonzero polynomial of degree at most two, verified by direct calculation. However, its roots at $s=0,1$ must both have even multiplicity, since $D(u,v_1)$ is nonnegative in a neighborhood of each overlap. This is impossible for a nonzero polynomial of degree at most two. Consequently, $D(u,v_1)$ must vanish for every $u \in [-1,1]$. In particular, $D(\pm 1,v_1)=0$. This implies that a local strategy with $A_1=\pm A_0$ also attains the quantum bound, contradicting the assumption of quantum advantage. The case $u_2=u_1$ but $v_2\neq v_1$ follows analogously. Hence, $(u_2,v_2)=(u_1,v_1)$.

Finally, for the dimension of the eigenspace, suppose by contradiction that the eigenspace had dimension at least two. However, any subspace of $\C^2 \ten \C^2$ with dimension larger than or equal to $2$ contains a product vector~\cite{Parthasarathy2004}. This implies that the eigenspace, which is a subspace of $\C^2 \ten \C^2$, contains a product vector, contradicting the assumption that $\vb$ has quantum advantage.
\end{proof}

\subsection{Proof of \Cref{thm:main}}

\begin{proof}
    We prove that there can be only one extremal correlation that maximizes a Tsirelson inequality with quantum advantage.
    Suppose that there are two nonlocal extremal correlations $\vP_1$ and $\vP_2$ that maximize the same Tsirelson inequality with quantum advantage. By~\citep[Proposition A.1]{Barizien25_CHSH_selftests}, they are realized by two-qubit strategies
    \begin{equation*}
    \bigl(\ket{\psi},\{A_x\},\{B_y\}\bigr),
    \quad
    \bigl(\ket{\psi'},\{A'_x\},\{B'_y\}\bigr).
    \end{equation*}
    Since both strategies attain the quantum bound of $\vb$, \cref{lem:overlap-unique} guarantees that the overlaps of the two strategies are equal.
    Consequently, there exist local unitaries $U_A,U_B,U'_A,U'_B$ such that
    \begin{equation*}
    U_AA_xU_A^\dagger=U'_AA'_xU_A^{\prime\dagger}=\bar A_x,\qquad
    U_BB_yU_B^\dagger=U'_BB'_yU_B^{\prime\dagger}=\bar B_y,
    \end{equation*}
    where $\bar A_x, \bar B_y$ may be chosen as
    \begin{equation*}
    \begin{aligned}
    \bar A_0&\coloneqq Z,&
    \bar A_1&\coloneqq u Z+\sqrt{1-u^2}\,X,\\
    \bar B_0&\coloneqq Z,&
    \bar B_1&\coloneqq v Z+\sqrt{1-v^2}\,X.
    \end{aligned}
    \end{equation*}
    These observables are used to construct the Bell operator $\bar S(u,v)$.
    Then, the two states, $\ket{\bar\psi}\coloneqq(U_A\otimes U_B)\ket{\psi}$ and $\ket{\bar\psi'}\coloneqq(U'_A\otimes U'_B)\ket{\psi'}$, are both eigenstates of $\bar S(u,v)$ associated with the quantum bound.
    \begin{equation*}
    \bar S(u,v)\ket{\bar\psi} = \ket{\bar\psi},\qquad \bar S(u,v)\ket{\bar\psi'} = \ket{\bar\psi'}.
    \end{equation*}
    By~\cref{lem:overlap-unique}, the eigenspace associated with the quantum bound of $\bar S(u,v)$ is one-dimensional.
    It follows that $\ket{\bar\psi'}=e^{i\varphi}\ket{\bar\psi}$ for some $\varphi\in\mathbb R$. 
    Therefore, for all $x,y$,
    \begin{equation*}
    \bra{\psi'}A'_x\otimes\openone\ket{\psi'}=
    \bra{\psi}A_x\otimes\openone\ket{\psi}, \quad
    \bra{\psi'}\openone\otimes B'_y\ket{\psi'}=
    \bra{\psi}\openone\otimes B_y\ket{\psi}, \quad
    \bra{\psi'}A'_x\otimes B'_y\ket{\psi'}=
    \bra{\psi}A_x\otimes B_y\ket{\psi}.
    \end{equation*}
    Thus, the two strategies generate the same correlations. 
    Therefore $\vP_1=\vP_2$.

    Next, we show that such a correlation is exposed. Assume by contradiction that there exists a nonlocal extremal and non-exposed correlation $\vP$ that maximizes a Tsirelson inequality $\vb$ with quantum advantage.
    However, the above shows that such $\vb$ can only be maximized by $\vP$ itself, so $F_{\vb}=\{\vP\}$. Therefore, $\vb$ can be used to expose $\vP$, contradicting the assumption that $\vP$ is non-exposed.
\end{proof}

\newpage

\section{The Dimension of Orthogonal Faces} \label{app:dimensionbound}

For a given $\vP$ obtained by a qubit realization as in Eq.~\eqref{eq:realizqubit}, we consider the five tangent vectors to $\vP$ among small variations of the measurements and state parameters:
\begin{equation}
\label{eq:TangentVec}
    \begin{aligned}
    \p_\theta &\vP \coloneqq \dfrac{\partial \vP}{\partial \theta} &&=
    -2 
    \l(\begin{array}{c|c|c}
        & c_{b_0}s_{2\theta} 
        & c_{b_1}s_{2\theta} \\ \hline
        c_{a_0}s_{2\theta} 
        & -s_{a_0}s_{b_0}c_{2\theta} 
        & -s_{a_0}s_{b_1}c_{2\theta} \\ \hline
        c_{a_1}s_{2\theta}
        & -s_{a_1}s_{b_0}c_{2\theta} 
        & -s_{a_1}s_{b_1}c_{2\theta}
    \end{array}\r)\\
    \p_{a_0} &\vP \coloneqq \dfrac{\partial \vP}{\partial a_0} &&=
    \l(\begin{array}{c|c|c}
        & 0 & 0 \\ \hline
        -s_{a_0}c_{2\theta}
        & c_{a_0}s_{b_0}s_{2\theta}-s_{a_0}c_{b_0}
        & c_{a_0}s_{b_1}s_{2\theta}-s_{a_0}c_{b_1} \\ \hline
        0 
        & 0 
        & 0
    \end{array}\r)\\
    \p_{a_1} &\vP \coloneqq \dfrac{\partial \vP}{\partial a_1} &&=
    \l(\begin{array}{c|c|c}
        & 0 & 0 \\ \hline
        0
        & 0
        & 0 \\ \hline
        -s_{a_1}c_{2\theta}
        & c_{a_1}s_{b_0}s_{2\theta}-s_{a_1}c_{b_0}
        & c_{a_1}s_{b_1}s_{2\theta}-s_{a_1}c_{b_1}
    \end{array}\r)\\
    \p_{b_0} &\vP \coloneqq \dfrac{\partial \vP}{\partial b_0} &&=
    \l(\begin{array}{c|c|c}
        & -s_{b_0}c_{2\theta} & 0\\ \hline
        0
        & s_{a_0}c_{b_0}s_{2\theta}-c_{a_0}s_{b_0}
        & 0\\ \hline
        0
        & s_{a_1}c_{b_0}s_{2\theta}-c_{a_1}s_{b_0}
        & 0
    \end{array}\r)\\
    \p_{b_1} &\vP \coloneqq \dfrac{\partial \vP}{\partial b_1} &&=
    \l(\begin{array}{c|c|c}
        & 0 & -s_{b_1}c_{2\theta}\\ \hline
        0
        & 0
        & s_{a_0}c_{b_1}s_{2\theta}-c_{a_0}s_{b_1}\\ \hline
        0
        & 0
        & s_{a_1}c_{b_1}s_{2\theta}-c_{a_1}s_{b_1}
    \end{array}\r)
    \end{aligned}
\end{equation}
The subspace spanned by these five vectors is denoted by
\begin{equation}
T_{\vP} \coloneqq \text{span} \Big(
    \p_\theta\vP, \p_{a_0}\vP, \p_{a_1}\vP, \p_{b_0}\vP, \p_{b_1}\vP 
    \Big).
\end{equation}
Every Bell functional $\vb$ maximized by $\vP$ is orthogonal to $T_{\vP}$. Indeed, every choice of the parameters gives a quantum correlation, so $\vb\cdot\vP(\theta',a_0',a_1',b_0',b_1')\leq1$ for all parameters, with equality at $(\theta,a_0,a_1,b_0,b_1)$. The gradient of this function vanishes there,
\begin{equation}\label{eq:first-order}
  \vb\cdot\p_i\vP=0, \qquad i=\theta,a_0,a_1,b_0,b_1 .
\end{equation}
We will first show that this is a five-dimensional subspace as long as $\vP$ is nonlocal, extremal, and does not arise from a maximally entangled state.
For an alternative proof of linear independence, see~\citep[Section IV.2]{Barizien24_PhD}.
Note that a relabeling maps $T_{\vP}$ linearly onto the tangent space of the relabeled point, so it preserves $\dim(T_{\vP})$. By~\cref{prop:FA}, we can therefore assume that the parameters satisfy~\cref{eq:extremal-corr}.
\begin{lemma}\label{lem:dim5}
    Let $\vP(\theta, a_0, a_1, b_0, b_1)$ be a nonlocal extremal correlation with parameters being inside the range specified in~\cref{eq:range}.
    For any $\theta \neq \pi/4$, $\dim (T_{\vP}) = 5$.
\end{lemma}
\begin{proof}

Suppose that the five vectors are not linearly independent, so that
\begin{equation}
\label{eq:LD}
    \lambda_{\theta}\p_\theta \vP + \lambda_{a_0} \p_{a_0} \vP + \lambda_{a_1} \p_{a_1} \vP + \lambda_{b_0} \p_{b_0} \vP + \lambda_{b_1} \p_{b_1} \vP = 0,
\end{equation}
with $\lambda_i$s not being all zeros, where $i=\theta, a_0, a_1, b_0, b_1$.
\begin{enumerate}
\item 
We first treat the cases with $a_0,b_0>0$.

Every single entry of the vector in~\cref{eq:LD} gives a linear equation in terms of $\lambda_\theta, \lambda_{a_0}, \lambda_{a_1}, \lambda_{b_0}, \lambda_{b_1}$.
The marginal terms are equivalent to the following equalities:
\begin{equation}
    \lambda_\theta \partial_\theta \braket{A_x} + 
    \lambda_{a_x}\partial_{a_x} \braket{A_x} = 0, \quad
    \lambda_\theta \partial_\theta \braket{B_y} +
    \lambda_{b_y} \partial_{b_y} \braket{B_y} = 0,
\end{equation}
With this, we obtain relations between $\lambda_\theta$ and $\lambda_i$ for $i = a_0, a_1, b_0, b_1$:
\begin{equation}
\label{eq:lbd-solu}
    \lambda_{i} = -2 \lambda_\theta \cot(i)\tan(2\theta), \quad i = a_0, a_1, b_0, b_1.
\end{equation}
The coefficients are well-defined since we are assuming $\theta \neq \pi/4$ and $a_0,b_0 \neq 0$.
Also, since the correlation is nonlocal, we have $a_1 > a_0 > 0$ and $b_1 > b_0 > 0$.
Plug the coefficients in~\cref{eq:lbd-solu} back to the correlator terms of~\cref{eq:LD} and get:
\begin{equation}
    \lambda_\theta \p_\theta \braket{A_xB_y} + 
    \lambda_{a_x}\p_{a_x} \braket{A_xB_y} +  
    \lambda_{b_y}\p_{b_y} \braket{A_xB_y} = 0.
\end{equation}
One finds that the following must be satisfied in order for the linear dependence relation to hold:
\begin{equation}
\label{eq:LD-half-angle}
\lambda_\theta \l[ 
    \dfrac{8(\a_x\b_y+t)(\a_x t-\b_y)(\a_x -\b_y t)(1 + \a_x\b_y t)}{\a_x\b_y(1+\a_x^2)(1+\b_y^2)(t^4-1)} 
    \r] = 0,
\end{equation}
for all $(x,y)=(0,0),(0,1),(1,0),(1,1)$.
For a nonlocal extremal correlation, we have
$0<t<1$, $\beta_0<\beta_1$, and
$\alpha_0\leq\beta_0t$. Hence
$
\alpha_0-\beta_1t<0,
\alpha_0t-\beta_1<0.
$
Since all remaining numerator factors in~\cref{eq:LD-half-angle} are strictly positive, and its denominator is nonzero, the equation corresponding to $(x,y)=(0,1)$ implies
$\lambda_\theta=0$.
By \cref{eq:lbd-solu}, all remaining coefficients also vanish, contradicting the assumed nontrivial linear dependence.

\item For the case with $0 = a_0 \leq b_0$ and $\theta \neq \pi/4$, pick the rows of~\cref{eq:LD} which correspond to the terms of $\braket{A_0}, \braket{A_1}, \braket{B_1}, \braket{A_0B_1}, \braket{A_1B_0}$.
Use the coefficients in front of the $\lambda_i$s to form the following $5 \times 5$ matrix:
\begin{equation}
M \coloneqq
\begin{pmatrix}
-2s_{2\theta} & 0 & 0 & 0 & 0 \\
-2c_{a_1}s_{2\theta} & 0 & -s_{a_1}c_{2\theta} & 0 & 0 \\
-2c_{b_1}s_{2\theta} & 0 & 0 & 0 & -s_{b_1}c_{2\theta} \\
0 & s_{b_1}s_{2\theta} & 0 & 0 & -s_{b_1} \\
2s_{a_1}s_{b_0}c_{2\theta} & 0
& c_{a_1}s_{b_0}s_{2\theta}-s_{a_1}c_{b_0}
& s_{a_1}c_{b_0}s_{2\theta}-c_{a_1}s_{b_0}
& 0
\end{pmatrix}.
\end{equation}
One can verify that the determinant of this matrix is 
\begin{equation}
    \det(M)=\dfrac{1}{2}\sin(a_1)\sin^2(b_1)\sin^2(4\theta)
    \Big[
        \sin(a_1)\cos(b_0)\sin(2\theta) - \cos(a_1)\sin(b_0)
    \Big].
\end{equation}
Nonlocality and the assumed parameter range give $\sin(a_1),\sin(b_1),\sin(4\theta) \neq 0$.
The term in brackets rewritten with the half-angle variables is proportional to
\begin{equation}
    2(\a_1t - \b_0)(1 + \a_1\b_0t) + \b_0(1+\a_1^2)(1-t^2).
\end{equation}
If $b_0 = 0$, the expression then reduces to $2\a_1t$, which cannot be zero since nonlocality requires $\a_1 > \a_0 = 0$ and $\theta > 0$.
If $b_0 > 0$, $\beta_0$, $1+\alpha_1^2$ and $1-t^2$ are strictly positive, and $\a_1t-\b_0\geq0$ by~\cref{eq:extremal-corr}, so the expression cannot be zero.
Hence, $\det(M) \neq 0$ when the considered correlation is nonlocal and extremal.
The five vectors are therefore linearly independent.
\end{enumerate}
\end{proof}

However, when $\theta = \pi/4$, the five tangent vectors are linearly {\em dependent}, since $\p_{a_0} \vP + \p_{a_1} \vP + \p_{b_0} \vP + \p_{b_1} \vP = 0$.
Nevertheless, by replacing one of these tangent vectors, we can still identify a five-dimensional subspace orthogonal to every Bell functional $\vb$ that is maximized by the correlation $\vP$.
To this end, define the vector of observables as
\begin{eqnarray}
\label{eq:obs-order}
    \vec{M} \coloneqq
    \begin{pmatrix}
        A_0, A_1, B_0, B_1, A_0B_0, 
        A_0B_1,
        A_1B_0,
        A_1B_1
    \end{pmatrix}
    \in \mathcal{L}(\hs_A \ten \hs_B)^8,
\end{eqnarray}
and the vector
\begin{equation}
\label{eq:vec-w}
    \vec{w} \coloneqq \bra{\psi} \vec{M} \ket{\phi_\theta}, \quad
    \ket{\psi} \coloneqq \cos{\theta} \ket{01} + \sin{\theta} \ket{10}. 
\end{equation}
Given that the correlation maximizes the inequality $\vb$, normalized to have quantum bound being one, the corresponding Bell operator $\hat S \coloneqq \vb \cdot \vec{M}$ satisfies $\hat S \ket{\phi_\theta} = \ket{\phi_\theta}$, since every state gives a quantum correlation, so the largest eigenvalue of $\hat S$ is at most one, and $\bra{\phi_\theta}\hat S\ket{\phi_\theta}=\vb\cdot\vP=1$.
Together with $\braket{\psi|\phi_\theta} = 0$, we arrive at $\vb \cdot \bra{\psi} \vec{M} \ket{\phi_\theta} = \vb \cdot \vec{w} = 0$, implying that $\vec{w}$ is orthogonal to $\vb$.
We now show that this additional vector, together with four tangent vectors, spans a five-dimensional subspace.
\begin{lemma}
Let $\vP$ be a nonlocal extremal correlation realized at $\theta=\pi/4$, with
$0\le a_0\le b_0\le a_1\le b_1<\pi$. Then
\[
\dim \Big( \text{span}
\left\{
\vec{w},\p_\theta\vP,
\p_{a_0}\vP,\p_{a_1}\vP,\p_{b_0}\vP
\right\} \Big)=5.
\]
\end{lemma}
\begin{proof}

    When $\theta = \pi/4$, the five vectors become
    \begin{equation}
    \Big(
    \vec{w},
    -\p_\theta\vP/2, 
    \p_{a_0}\vP, 
    \p_{a_1}\vP, 
    \p_{b_0}\vP
    \Big) = 
    \begin{pmatrix}
    s_{a0} & c_{a0} & 0 & 0 & 0 \\
    s_{a1} & c_{a1} & 0 & 0 & 0 \\
    s_{b0} & c_{b0} & 0 & 0 & 0 \\
    s_{b1} & c_{b1} & 0 & 0 & 0 \\
    0 & 0 & c_{a0}s_{b0}-s_{a0}c_{b0}
    & 0 & s_{a0}c_{b0}-c_{a0}s_{b0} \\
    0 & 0 & c_{a0}s_{b1}-s_{a0}c_{b1}
    & 0 & 0 \\
    0 & 0 & 0 & c_{a1}s_{b0}-s_{a1}c_{b0}
    & s_{a1}c_{b0}-c_{a1}s_{b0} \\
    0 & 0 & 0 & c_{a1}s_{b1}-s_{a1}c_{b1}
    & 0
    \end{pmatrix}
    \end{equation}
The first two columns are supported on the first four coordinates, whereas the remaining three columns are supported on the last four coordinates. It therefore suffices to establish linear independence within each group. The minor of the first two columns corresponding to the first and fourth rows is
\[
\begin{vmatrix}
s_{a0} & c_{a0}\\
s_{b1} & c_{b1}
\end{vmatrix}
=\sin(a_0-b_1)\neq 0,
\]
given that $a_0<b_1$ for a nonlocal correlation.
Thus $\vec{w}$ and $\p_\theta\vP$ are linearly independent.

For the last three columns, since $\p_{a_x} \vP$ are orthogonal to each other and nonzero (because $a_0<b_1$ and $b_0<b_1$), it suffices to show that $\p_{b_0}\vP$ lies outside of the span of the other two.
Assume that $\lambda_0\p_{a_0}\vP +\lambda_1\p_{a_1}\vP = \p_{b_0}\vP$.
Then, the entry that corresponds to $\braket{A_0B_1}$ gives
$\sin(a_0-b_1) = 0$ or $\lambda_0 = 0$.
Since nonlocal correlations require $a_0 < b_1$, $\lambda_0$ must be zero.
With that, the entry that corresponds to $\braket{A_0B_0}$ gives $a_0 = b_0$, and the entry that corresponds to $\braket{A_1B_1}$ gives $a_1 = b_1$ (if $a_1=b_0$, then $a_0=b_0=a_1$ and the correlation is local).
Thus, $\braket{A_0B_0} = \braket{A_1B_1} = 1$.
These also imply that $\braket{A_0B_1} = \cos(a_0 - b_1) = \cos(a_1 - b_0) = \braket{A_1B_0}$.
The correlation does not violate any CHSH inequality and is therefore local, contradicting our assumption.
(We note that this belongs to the Class 4b correlation family in~\cite{Chen23}.)
Hence, the last three vectors are mutually linearly independent.
Together with the independence of the first two vectors, this proves
that the five vectors span a five-dimensional subspace.
\end{proof}

All in all, whether a nonlocal extremal correlation $\vP$ is realized by a maximally entangled state or not, the inequality $\vb$ that it maximizes is always orthogonal to a five-dimensional subspace.
Together with the normalization, $\vb \cdot \vP = 1$, this implies that the affine dimension of the orthogonal face (see~\cref{eq:orth-face}) is at most~$2$. This proves~\cref{thm:dual}.

Remark that a different choice of vectors could have been used in the proof; see e.g.~\citep[Appendix D.1]{Barizien25_CHSH_selftests}. Actually, the vector $\vec{w}$ defined in~\cref{eq:vec-w} is one such kind, but many more can be obtained in a similar manner. To this end, one can define
\begin{equation}
\vec{T}_{\ket{\psi}} \coloneqq \bra{\psi} \vec{M} \ket{\phi_\theta},
\end{equation}
where $\ket{\psi}$ is chosen to be orthogonal to $\ket{\phi_\theta}$. With this, we see that $\vec w = \cos(\theta) \vec{T}_{\ket{01}} + \sin(\theta) \vec{T}_{\ket{10}}$. By a similar argument as for $\vec{w}$, we have $\vb \cdot \vec{T}_{\ket{\psi}} = 0$. We list a few examples of such vectors below, which can in fact be expressed as linear combinations of the tangent vectors of $\vP$.
\begin{equation}
\begin{aligned}
-\cos(2\theta)\,\vec T_{\ket{01}}
&=\sin\theta\bigl(\p_{a_0}\vP+\p_{a_1}\vP\bigr)
  +\cos\theta\bigl(\p_{b_0}\vP+\p_{b_1}\vP\bigr),\\
-\cos(2\theta)\,\vec T_{\ket{10}}
&=\cos\theta\bigl(\p_{a_0}\vP+\p_{a_1}\vP\bigr)
  +\sin\theta\bigl(\p_{b_0}\vP+\p_{b_1}\vP\bigr),\\
-2\vec T_{\ket{\psi_\theta}}
&=\p_\theta\vP,
\qquad
\ket{\psi_\theta}
\coloneqq\sin\theta\ket{00}-\cos\theta\ket{11}.
\end{aligned}
\end{equation}
Hence, when $\theta \neq \pi/4$,
these vectors lie in the span of the tangent vectors of $\vP$, $\p_\theta \vP, \p_{a_x}\vP, \p_{b_y}\vP$.

\newpage

\newcommand{\bbeta}{\vec{\beta}}
\newcommand{\PP}{\vec{P}}
\newcommand{\Pth}{\vec{P}_{\theta}}
\newcommand{\DD}{\vec{D}}
\newcommand{\fvec}[1]{\vec{#1}}
\newcommand{\bchsh}{\vec{\beta}_{\rm CHSH}}
\newcommand{\Lset}{\mathcal{L}}
\newcommand{\wl}{\omega_l}
\newcommand{\Poly}{\mathcal{P}}
\newcommand{\Adm}{\mathcal{A}}
\newcommand{\cth}{c_{2\theta}}
\newcommand{\sth}{s_{2\theta}}
\newcommand{\kth}{\kappa_{\theta}}
\newcommand{\detI}{\mathrm{I}}
\newcommand{\detII}{\mathrm{II}}
\newcommand{\LevA}{1{+}A{+}B{+}BB'{+}AB{+}ABB'}
\newcommand{\LevB}{1{+}A{+}B{+}AB{+}AA'B{+}ABB'}
\providecommand{\todo}[1]{\textcolor{red}{[#1]}}

\section{Orthogonal faces of partially entangled Tsirelson points}\label{app:fam-family}

Let us consider the family of partially entangled Tsirelson points defined by
\begin{equation}\label{eq:fam-P}
  \Pth \coloneqq \Big( \tfrac{\cth}{\sqrt2},\, \tfrac{\cth}{\sqrt2},\, \cth,\, 0,\,
  \tfrac{1}{\sqrt2},\, \tfrac{\sth}{\sqrt2},\, \tfrac{1}{\sqrt2},\, -\tfrac{\sth}{\sqrt2} \Big),
\end{equation}
which is written in the order of~\cref{eq:vec-order}, and obtained from a partially entangled state and observables
\begin{equation}\label{eq:real}
  \begin{aligned}
    |\phi_\theta\rangle &\coloneqq \cos\theta\,|00\rangle + \sin\theta\,|11\rangle,\\
    A_x &\coloneqq \frac{1}{\sqrt{2}}(\sigma_z+(-1)^x\sigma_x),\\
    B_0 &\coloneqq \sigma_z, \quad B_1 \coloneqq \sigma_x.
  \end{aligned}
\end{equation}
For $\theta=\pi/4$ this point is the Tsirelson point~\cite{Barizien24-extremal}. We check extremality with~\cref{prop:FA}. Exchanging the parties and flipping the outcome of $B_1$ brings the settings to $(a_0,b_0,a_1,b_1)=(0,\pi/4,\pi/2,3\pi/4)$, within the range~\cref{eq:range}. The half-angle variables~\cref{eq:half-angles} are then $\a_0=0$, $\b_0=\tan(\pi/8)$, $\a_1=1$ and $\b_1=\tan(3\pi/8)$, so the conditions~\cref{eq:extremal-corr} reduce to $\tan\theta\geq\tan(\pi/8)$. Hence $\Pth$ is extremal for $\pi/8\leq\theta\leq\pi/4$, with strict full alternation for $\theta>\pi/8$.
For $\pi/8<\theta\leq\pi/4$ it is also exposed. By \cref{thm:fam}, the inequalities $\bbeta_\theta(r_0,r_1)$ with $(r_0,r_1)$ in the interior of $\Poly_\theta$ lie in $\Pth^\perp$ and have local bound below $1$, so they have quantum advantage, and \cref{thm:main} applies.
At $\theta=\pi/8$, it violates strict full alternation, so \Cref{prop:not-strict-FA} ensures it is non-exposed. Note that some joint probabilities of $\vP_{\pi/8}$ vanish, for instance
\begin{equation}
  P_{\pi/8}(-1,-1|1,1) = \tfrac14\big(1 + \braket{A_1B_1} - \braket{A_1}-\braket{B_1}\big)\Big|_{\theta=\pi/8} = 0,
\end{equation}
so $\vec{P}_{\pi/8}$ lies on the boundary of the no-signaling set. 
For later convenience, we define the parameter
\begin{equation}\label{eq:fam-xi}
  \xi \coloneqq \frac{\tan\theta-(\sqrt2-1)}{2-\sqrt2},
\end{equation}
which maps $\theta\in(\pi/8,\pi/4]$ onto $\xi\in(0,1]$. It will be helpful to express quantities as rational functions of $\xi$ with coefficients in $\mathbb Q(\sqrt2)$.

\subsection{Plane of candidate Bell functionals}

Let us parametrize the Bell expressions with $\bbeta\cdot\Pth=1$ that are orthogonal to the tangent vectors of $\Pth$ with respect to variations in the parameters $\theta, a_x, b_y$, as in~\cref{eq:first-order}. For $\theta<\pi/4$, they form a two-dimensional plane by Appendix~\ref{app:dimensionbound}, and by construction it contains the orthogonal face of $\Pth$. In place of the variation of $\theta$, we use a stronger condition. Let $\bbeta\in\Pth^\perp$ and let $S$ be its Bell operator with the measurements~\eqref{eq:real}. Then
\begin{equation*}
  \lambda_{\max}(S) \le \omega_q(\bbeta) = \bbeta\cdot\Pth = \bra{\phi_\theta}S\ket{\phi_\theta} \le \lambda_{\max}(S),
\end{equation*}
so $\ket{\phi_\theta}$ is an eigenvector of $S$ with eigenvalue $1$. Together with the variations of the measurements, every $\bbeta\in\Pth^\perp$ satisfies
\begin{equation}\label{eq:fam-tangency}
  S\ket{\phi_\theta} = \ket{\phi_\theta}, \qquad
  \frac{\partial\,\bbeta\cdot\PP}{\partial a_x}\,\Big|_{\PP=\Pth}
  = \frac{\partial\,\bbeta\cdot\PP}{\partial b_y}\,\Big|_{\PP=\Pth} = 0,
  \quad x,y \in \{0,1\}.
\end{equation}

\begin{proposition}\label{prop:fam-plane}
  Let $\kth\coloneqq1+\sth^2$. For $0<\theta\leq\pi/4$ the solutions of Eq.~\eqref{eq:fam-tangency} form the affine plane
  \begin{equation}\label{eq:fam-plane}
    \bbeta_\theta(r_0,r_1) \coloneqq \bbeta^{(0)} + r_0\,\bbeta^{(1)} + r_1\,\bbeta^{(2)},
    \qquad (r_0,r_1) \in \mathbb R^2,
  \end{equation}
  with
  \begin{equation}\label{eq:fam-basis}
    \begin{aligned}
      \bbeta^{(0)} &\coloneqq \tfrac{1}{\kth}\Big(\cth\,B_0 + \tfrac{\sth^2}{\sqrt2}\,(A_0+A_1)B_0 + \tfrac{\sth}{\sqrt2}\,(A_0-A_1)B_1\Big),\\
      \bbeta^{(1)} &\coloneqq \tfrac{1}{\sqrt2}\,(A_0+A_1) - \tfrac{1}{\kth}\Big(2\sth^2\,B_0 + \tfrac{\cth}{\sqrt2}\,(A_0+A_1)B_0 - \tfrac{\sth\cth}{\sqrt2}\,(A_0-A_1)B_1\Big),\\
      \bbeta^{(2)} &\coloneqq \tfrac{\sth}{\sqrt2}\,(A_0-A_1) - B_1 + \tfrac{\cth}{\sqrt2}\,(A_0+A_1)B_1 .
    \end{aligned}
  \end{equation}
\end{proposition}
\noindent Note that at $\theta=\pi/4$, \cref{eq:fam-basis} reproduces the family of Bell expressions given in Ref.~\cite{Barizien24-extremal},
  \begin{equation}
      \bbeta_{\pi/4}(r_0,r_1) = \bchsh + r_0\,\Big(\tfrac{A_0+A_1}{\sqrt2} - B_0\Big) + r_1\,\Big(\tfrac{A_0-A_1}{\sqrt2} - B_1\Big),
  \end{equation}
where $\bchsh\cdot\PP\coloneqq \tfrac{1}{2\sqrt2}\,(\braket{A_0B_0}+\braket{A_0B_1}+\braket{A_1B_0}-\braket{A_1B_1})$ is the CHSH expression normalized to quantum bound $1$. 
Its local bound is $1/\sqrt2<\bchsh\cdot\Pth=(1+\sth)/2$.

\begin{proof}
  Following the method of Ref.~\cite{Barizien24-extremal}, we define the qubit operators built from $\openone$, $A_x$, $B_y$ and $A_xB_y$ that annihilate $\ket{\phi_\theta}$, spanned by
  \begin{equation}
    \begin{aligned}
      N_1 &\coloneqq \tfrac{1}{\sqrt2}\,(A_0+A_1) - B_0,\\
      N_2 &\coloneqq \openone - \tfrac{1}{\sqrt2}\,(A_0+A_1)B_0,\\
      N_3 &\coloneqq \tfrac{1}{\sqrt2}\,\big((A_0-A_1)B_0 + (A_0+A_1)B_1\big) - \tfrac{1-\sth}{\cth}\,\Big(\tfrac{A_0-A_1}{\sqrt2} + B_1\Big),\\
      N_4 &\coloneqq (1+\sth)\Big(\tfrac{A_0-A_1}{\sqrt2} - B_1\Big) - \tfrac{\cth}{\sqrt2}\,\big((A_0-A_1)B_0 - (A_0+A_1)B_1\big),\\
      N_5 &\coloneqq \openone - \tfrac{\sth}{\sqrt2}\,(A_0-A_1)B_1 - \cth\,B_0 .
    \end{aligned}
  \end{equation}
  The first condition in Eq.~\eqref{eq:fam-tangency} then holds if and only if
  \begin{equation}
    S = \openone + \sum_{i=1}^5 p_i\,N_i
  \end{equation}
  for some real $p_i$. Setting the identity component of the Bell operator to zero (corresponding to a shift in the Bell value of the fully mixed correlation) imposes
  \begin{equation}
    1 + p_2 + p_5 = 0 .
  \end{equation}
  Stationarity in the four measurement angles from \cref{eq:fam-tangency} gives two further independent conditions,
  \begin{equation}
    \cth\,p_1 - p_2 + \sth^2\,p_5 = 0, \qquad p_3 = \cth\,p_4 .
  \end{equation}
  Solving for $p_2$, $p_3$, $p_5$ and setting $r_0\coloneqq p_1$, $r_1\coloneqq2p_4$ gives the plane~\eqref{eq:fam-plane} with the basis~\eqref{eq:fam-basis}. The case $\theta=\pi/4$ needs to be treated specially since $N_3$ is singular then, but the result holds from Ref.~\cite{Barizien24-extremal}.
\end{proof}

\begin{remark}[Mirror symmetry]\label{rem:fam-mirror}
  The relabeling $A_0\leftrightarrow A_1$, $B_1\mapsto-B_1$ is a conjugation by $\sigma_z\otimes\sigma_z$,
  \begin{equation}
    (\sigma_z\otimes\sigma_z)\,A_0\,(\sigma_z\otimes\sigma_z) = A_1, \qquad
    (\sigma_z\otimes\sigma_z)\,B_1\,(\sigma_z\otimes\sigma_z) = -B_1,
  \end{equation}
  and it fixes $B_0$ and $\ket{\phi_\theta}$. Therefore, this relabeling fixes $A_0+A_1$, $B_0$, $(A_0+A_1)B_0$ and $(A_0-A_1)B_1$, and flips $A_0-A_1$, $B_1$ and $(A_0+A_1)B_1$, and so it also fixes $\bbeta^{(0)}$ and $\bbeta^{(1)}$ and flips $\bbeta^{(2)}$. Hence it maps $\bbeta_\theta(r_0,r_1)$ to $\bbeta_\theta(r_0,-r_1)$, and all sets in the $(r_0,r_1)$ plane below are symmetric under $r_1\mapsto-r_1$. We write $V'$ for the mirror image of a point $V$.
\end{remark}

\subsection{Local constraints in the plane}

Not every candidate in the two-dimensional plane defined by \cref{eq:fam-basis} is a Tsirelson inequality. In order to exclude some candidates from this set, we consider the dual local set $\Lset^\ast\coloneqq\{\bbeta:\bbeta\cdot\PP\le1\ \forall\,\PP\in\Lset\}$. Since $\mathcal{Q}^* \subset \mathcal{L}^*$, every $\vb$ in the orthogonal face has local bound at most $1$. Since $\Lset$ is the convex hull of the sixteen deterministic strategies $\DD$, it suffices to ensure
\begin{equation}\label{eq:fam-local}
  \bbeta_\theta(r_0,r_1)\cdot\DD \le 1 \qquad \forall\,\DD\in\Ext(\Lset).
\end{equation}
These half-planes identify the convex polygon
\begin{equation}
    \Poly_\theta \coloneqq \big\{ (r_0,r_1) :\ \bbeta_\theta(r_0,r_1)\cdot\DD \le 1\ \ \forall\,\DD\in\Ext(\Lset) \big\}
\end{equation}
as a subregion of the plane of candidates. 

Only the eight deterministic strategies with $\bchsh\cdot\DD=1/\sqrt{2}$ give active constraints (condition (b) in the proof of \cref{prop:fam-polygon}). For $\epsilon,\epsilon'\in\{\pm1\}$ we label them
\begin{equation}
  \begin{aligned}
    \detI(\epsilon,\epsilon')  &:\quad A_0 = A_1 = B_0 = \epsilon,\quad B_1 = \epsilon',\\
    \detII(\epsilon,\epsilon') &:\quad A_0 = -A_1 = B_1 = \epsilon,\quad B_0 = \epsilon' .
  \end{aligned}
\end{equation}
Each strategy gives rise to a line in the plane, which identifies the Bell expressions that it maximizes together with $\Pth$. The constraints of $\detI(\epsilon,\epsilon')$ and $\detII(\epsilon,\epsilon')$ are the half-planes
\begin{subequations}\label{eq:fam-lines}
  \begin{align}
    \eta_\epsilon\,r_0 + \epsilon'\kth\big(\sqrt2\,\epsilon\cth-1\big)\,r_1
      &\le \kth - \epsilon\cth - \sqrt2\,\sth^2,\\
    \sth\big(\sqrt2\,\cth-2\epsilon'\sth\big)\,r_0 + \epsilon\kth\big(\sqrt2\,\sth-1\big)\,r_1
      &\le \kth - \epsilon'\cth - \sqrt2\,\sth,
  \end{align}
\end{subequations}
where $\eta_\epsilon\coloneqq\epsilon\big(\sqrt2-(2-\sqrt2)\sth^2\big)-\sqrt2\,\cth$ and $\kth\coloneqq1+\sth^2$.

\begin{figure}[b]
  \centering
  \includegraphics[width=0.9\textwidth]{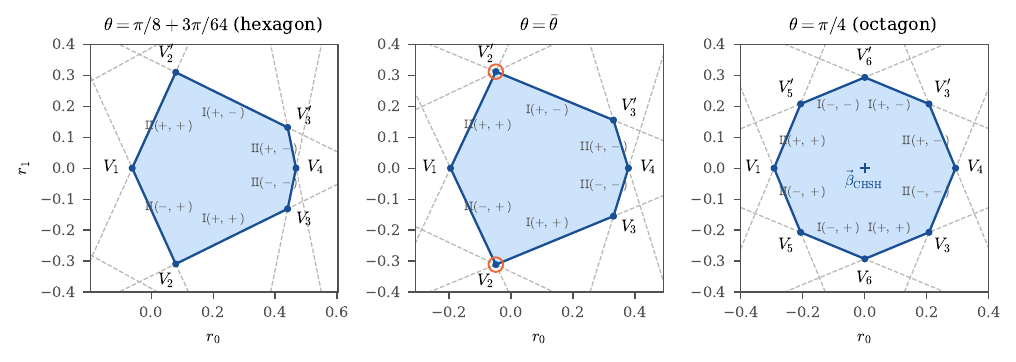}
  \caption{The local polygon $\Poly_\theta$ below $\bar\theta$, at $\bar\theta$ and at $\theta=\pi/4$, with the vertices of Proposition~\ref{prop:fam-polygon}. Each edge is labeled by the deterministic point whose line carries it. The dashed lines are the lines of the eight deterministic points with $\bchsh\cdot\DD=1/\sqrt{2}$. At $\bar\theta$ the lines of $\detI(+,+)$, $\detII(-,+)$ and $\detI(-,+)$ meet at the circled vertex $V_2$, and their mirror images meet at $V_2'$.}
  \label{fig:fam-vertices}
\end{figure}

\begin{proposition}[Local polygon]\label{prop:fam-polygon}
  Let $\theta\in(\pi/8,\pi/4]$. Define $\bar\theta\coloneqq\tfrac12\arcsin\bar s\simeq0.6445$, where $\bar s$ is the root of $2s^3+1=\sqrt2\,(1+s)$ in $(1/\sqrt2,1)$.
  We write $\ell\cap\ell'$ for the intersection of the boundary lines of two constraints~\eqref{eq:fam-lines}.
  \begin{enumerate}
    \item[(i)] For $\pi/8<\theta<\bar\theta$, $\Poly_\theta$ is a hexagon with vertices $V_1,V_2,V_3,V_4,V_3',V_2'$, in this order, where
      \begin{equation}
        \begin{aligned}
          V_1 &\coloneqq \detII(+,+)\cap\detII(-,+), &\qquad V_2 &\coloneqq \detI(+,+)\cap\detII(-,+),\\
          V_3 &\coloneqq \detI(+,+)\cap\detII(-,-),  &\qquad V_4 &\coloneqq \detII(+,-)\cap\detII(-,-).
        \end{aligned}
      \end{equation}
    \item[(ii)] For $\bar\theta<\theta\le\pi/4$, $\Poly_\theta$ is an octagon with vertices $V_1,V_5,V_6,V_3,V_4,V_3',V_6',V_5'$. The vertex $V_2$ is replaced by
      \begin{equation}
        V_5 \coloneqq \detII(-,+)\cap\detI(-,+), \qquad V_6 \coloneqq \detI(+,+)\cap\detI(-,+).
      \end{equation}
    \item[(iii)] At $\theta=\bar\theta$ the lines of $\detI(+,+)$, $\detII(-,+)$ and $\detI(-,+)$ are concurrent, and so are their mirror images. $\Poly_{\bar\theta}$ is a hexagon with two vertices where three constraints are active.
    \item[(iv)] The vertices of $\Poly_\theta$ on the $r_0$ axis are
      \begin{equation}
        V_1 = \Big( \frac{\kth-\cth-\sqrt2\,\sth}{\sth(\sqrt2\,\cth-2\sth)},\, 0 \Big), \qquad
        V_4 = \Big( \frac{\kth+\cth-\sqrt2\,\sth}{\sth(\sqrt2\,\cth+2\sth)},\, 0 \Big).
      \end{equation}
      At $\theta=\pi/4$, $\Poly_\theta$ is the regular octagon of circumradius $1-1/\sqrt2$~\cite{Barizien24-extremal}.
    \item[(v)] As $\theta\to\pi/8$, $\Poly_\theta$ converges to the segment $r_0=1/\sqrt2$, $|r_1|\le\sqrt2-1$. At $\theta=\pi/8$ itself, the feasible set is the longer segment $r_0=1/\sqrt2$, $|r_1|\le1$.
  \end{enumerate}
\end{proposition}
The first three cases are illustrated in Fig.~\ref{fig:fam-vertices}. This proposition allows us to obtain an outer bound on the orthogonal face $\vP_\theta^\perp$ in the two-dimensional plane identified in \Cref{prop:fam-plane}. In the following subsection, we actually prove that this outer approximation is tight.

\begin{proof}
  Write each constraint~\eqref{eq:fam-lines} as $\ell_{\DD}(r_0,r_1)\le b_{\DD}$, with $\ell_{\DD}$ linear. Its boundary line is $\ell_{\DD}(r_0, r_1)=b_{\DD}$. A point of a convex polygon is a vertex if and only if it lies in the polygon and two independent constraints are active there. Each listed point $V$ is the intersection of two lines, so its coordinates solve a $2\times2$ linear system. The list in (i) and (ii) is therefore the list of vertices as soon as, for every $\theta$ in the range,
  \begin{itemize}
    \item[(a)] the two lines through each $V$ are not parallel,
    \item[(b)] the other fourteen constraints are strict at each $V$, that is $\ell_{\DD}(V) < b_{\DD}$.
    \item[(c)] two consecutive points of the list lie on a common line.
  \end{itemize}
  By (a) and (b), every listed point is a vertex. By (c), consecutive vertices are joined by an edge. These edges close into a cycle, which is the whole boundary, so there is no other vertex. Condition (c) holds by construction, as the labels in (i) and (ii) show. For instance $V_1$ and $V_2$ both lie on the line of $\detII(-,+)$.

  Notice that $\tan\theta$ is affine in $\xi$, and
  \begin{equation}
    \cth = \frac{1-\tan^2\theta}{1+\tan^2\theta}, \qquad \sth = \frac{2\tan\theta}{1+\tan^2\theta}.
  \end{equation}
  So the determinant in (a) and every condition in (b) is a rational function $R(\xi)$ with coefficients in $\mathbb Q(\sqrt2)$. Each condition thus becomes a statement of the form
  \begin{equation}
    R(\xi) > 0 \qquad \forall\,\xi\in I,
  \end{equation}
  with $I=(0,\bar\xi)$ for the hexagon and $I=(\bar\xi,1]$ for the octagon, up to the sign of $R$. This statement holds if and only if the numerator and the denominator of $R$ have no roots in $I$, and if $R$ is positive at one point of $I$. We check these statements exactly with Mathematica~\cite{Code}.

  The transition happens when the lines of $\detI(+,+)$, $\detII(-,+)$ and $\detI(-,+)$ meet at one point. Three lines $a_ir_0+b_ir_1=c_i$, not all parallel, do so if and only if
  \begin{equation}
    \det\begin{pmatrix} a_1 & b_1 & c_1\\ a_2 & b_2 & c_2\\ a_3 & b_3 & c_3 \end{pmatrix} = 0 .
  \end{equation}
  Here the determinant is a polynomial in $(\cth,\sth)$. Eliminating $\cth$ with $\cth^2+\sth^2=1$ gives, up to a constant,
  \begin{equation}
    s\,\big(s-\tfrac{1}{\sqrt2}\big)^2\,(1+s^2)^4\,\big(2s^3-\sqrt2\,s+1-\sqrt2\big), \qquad s\coloneqq \sth .
  \end{equation}
  The factors $s$ and $1+s^2$ do not vanish on the range. The factor $(s-1/\sqrt2)^2$ vanishes at $\theta=\pi/8$, where the polygon collapses. So the transition is the root $\bar s$ of the last factor, which is the equation $2s^3+1=\sqrt2\,(1+s)$. In terms of $\xi$, $\bar\theta$ corresponds to $\bar\xi\simeq0.5759953$. At $\bar\xi$ exactly one further $\ell_{\DD}(V)\le b_{\DD}$ constraint is saturated, at $V_2$ and at $V_2'$. This gives item (iii).

  $V_1$ and $V_4$ are their own mirror images, so they lie on the $r_0$ axis. Setting $r_1=0$ in either of their two lines gives $r_0$. Item (v) follows by letting $\xi\to0$ in the vertex coordinates, whose denominators do not vanish at $\xi=0$. At $\theta=\pi/8$ we have $\cth=\sth=1/\sqrt2$ and $\kth=3/2$, and the constraints of $\detI(+,+)$ and $\detII(+,+)$ become
\begin{equation*}
  (3\sqrt2-4)\,r_0 \le 3-2\sqrt2, \qquad (2-\sqrt2)\,r_0 \ge \sqrt2-1 .
\end{equation*}
So $r_0=1/\sqrt2$. On this line every other constraint either holds with equality or reads $\pm r_1\le1$, so $\Poly_{\pi/8}$ is the segment $|r_1|\le1$. Its endpoints $\bbeta_{\pi/8}(1/\sqrt2,\pm1)$ are the positivity constraints $P(-1,+1|0,1)\ge0$ and $P(-1,-1|1,1)\ge0$, which $\vP_{\pi/8}$ saturates. They lie in $\Q^*$, so by convexity the orthogonal face of $\vP_{\pi/8}$ is the whole segment. Every point of this segment is also maximized by the deterministic point $A_0=A_1=B_0=B_1=1$, so $\vP_{\pi/8}$ is not exposed.
\end{proof}

\subsection{Sum-of-squares certificates}\label{app:fam-sos}

In this subsection, we certify that the quantum bound of the inequalities corresponding to the vertices of $\Poly_\theta$ is $1$. To do so, we consider $\fvec{M}$, the vector of monomials of an NPA level~\cite{NPA08}. A sum-of-squares (SOS) certificate of $\bbeta$ is an identity
\begin{equation}
  \openone - S_{\bbeta} = \fvec{M}^\dagger\, W\, \fvec{M}, \qquad W \succeq 0,
\end{equation}
of noncommutative polynomials modulo $A_x^2=B_y^2=\openone$ and $[A_x,B_y]=0$, where $S_{\bbeta}$ is the formal polynomial associated to $\vb$~\cite{Barizien24-extremal}. Writing $W=L^{\mathsf T}L$ gives $\openone-S_{\bbeta}=\sum_kP_k^\dagger P_k$ with $P_k\coloneqq\sum_iL_{ki}M_i$. So $S_{\bbeta}\preceq\openone$ for all observables, independently of the dimension of the Hilbert space.

\begin{proposition}[Sum-of-squares certificates]\label{thm:fam-sos}
  For every $\theta\in(\pi/8,\pi/4]$ and every vertex $V$ of $\Poly_\theta$ there is a matrix $W_\theta\succeq0$ such that
  \begin{equation}\label{eq:fam-sos}
    \openone - S_{\bbeta_\theta(V)} = \fvec{M}^\dagger\, W_\theta\, \fvec{M}
  \end{equation}
  modulo $A_x^2=B_y^2=\openone$ and $[A_x,B_y]=0$. For three vertex families the vector $\fvec{M}$ contains the monomials of the NPA level $\LevA$, and for the other three those of $\LevB$. The entries of $W_\theta$ are rational functions of $\tan\theta$ with coefficients in $\mathbb Q(\sqrt2)$.
\end{proposition}

\begin{proof}
    For a vertex inequality $\bbeta=\bbeta_\theta(V)$ we have $\bra{\phi_\theta}S_{\bbeta}\ket{\phi_\theta}=1$, so the SOS gives $\sum_k\|P_k\ket{\phi_\theta}\|^2=0$, and every $P_k$ annihilates $\ket{\phi_\theta}$.\footnote{Here, we slightly abuse notation. Formally speaking, when writing $P_k\ket{\phi_\theta}$, the object $P_k$  is assumed to be the evaluation of the formal polynomial $P_k$ on the measurements corresponding to the qubit realization of $\vP_\theta$.} In the same way, every $P_k$ vanishes at the saturated deterministic strategies. The polynomials in the span of $\fvec M$ with these two properties form a subspace. Let $\fvec{N}\coloneqq(N_1,\dots,N_r)$ be a basis of it, which depends on $\theta$ and $\bbeta$. It is then sufficient to look for a certificate of the form
    \begin{equation}\label{eq:fam-reduction}
      \openone - S_{\bbeta} = \fvec{N}^\dagger\, W'\, \fvec{N}, \qquad W' \succeq 0.
    \end{equation}  

    For each vertex family, we exhibit a single certificate valid on its whole range of $\xi$, given in Table~\ref{tab:fam-sos}. It has the form
    \begin{equation}
      \mu(\xi)\big(\openone-S_{\bbeta_\theta(V)}\big) = (\fvec N)^\dagger W'(\xi)\, (\fvec{N}),
    \end{equation}
    where $\mu(\xi)>0$ is the common denominator of the coefficients of $S_{\bbeta_\theta(V)}$, and $W'(\xi)$ is a symmetric matrix whose entries are polynomials in $\xi$ with coefficients in $\mathbb Q(\sqrt2)$. Note that the matrix $W'$ is $9\times9$ or $11\times11$, with polynomial entries of degree at most five (Table~\ref{tab:fam-sos}). We use the larger level for the three vertex families that reach the transition at $\bar\theta$. At $\bar\theta$ the smaller level admits no certificate for the triple-point inequality. All these results were first observed numerically with the NPA hierarchy~\cite{Andrea}, and Table~\ref{tab:fam-npa} compares the performance of each level. Note that we did not investigate whether these certificates are the smallest ones, in the size of $W'$, in the degree of its entries or in the NPA level. We only claim that each of them is a valid certificate.

    We verified the SOS exactly, modulo $A_x^2=B_y^2=\openone$ and $[A_x,B_y]=0$, with Mathematica and the NCAlgebra package~\cite{NCAlgebra}. The leading principal minors of $W'(\xi)$ are polynomials in $\xi$, and \texttt{Resolve} proves them positive on the range, except at $\bar\xi$ for the three families that reach it. So $W'(\xi)\succ0$ by Sylvester's criterion. At $\bar\xi$, $W'$ is a limit of positive definite matrices, so $W'\succeq0$. The mirror vertices follow from Remark~\ref{rem:fam-mirror}. The full matrices of all six families are given in the code repository~\cite{Code}.
\end{proof}

\begin{table}[t]
  \begin{ruledtabular}
    \begin{tabular}{llllll}
      Vertex        & Saturated strategies            & $(A_0,A_1,B_0,B_1)$             & Range               & NPA Level   & $W'$\\
      \hline
      $V_1$         & $\detII(+,+)$, $\detII(-,+)$   & $(1,-1,1,1)$, $(-1,1,1,-1)$     & $\xi\in(0,1]$       & $\LevA$ & $9\times9$, degree $3$\\
      $V_4$         & $\detII(+,-)$, $\detII(-,-)$   & $(1,-1,-1,1)$, $(-1,1,-1,-1)$   & $\xi\in(0,1]$       & $\LevA$ & $9\times9$, degree $3$\\
      $V_3$, $V_3'$ & $\detI(+,+)$, $\detII(-,-)$    & $(1,1,1,1)$, $(-1,1,-1,-1)$     & $\xi\in(0,1]$       & $\LevA$ & $9\times9$, degree $5$\\
      $V_2$, $V_2'$ & $\detI(+,+)$, $\detII(-,+)$    & $(1,1,1,1)$, $(-1,1,1,-1)$      & $\xi\in(0,\bar\xi]$ & $\LevB$ & $11\times11$, degree $4$\\
      $V_5$, $V_5'$ & $\detII(-,+)$, $\detI(-,+)$    & $(-1,1,1,-1)$, $(-1,-1,-1,1)$   & $\xi\in[\bar\xi,1]$ & $\LevB$ & $11\times11$, degree $4$\\
      $V_6$, $V_6'$ & $\detI(+,+)$, $\detI(-,+)$     & $(1,1,1,1)$, $(-1,-1,-1,1)$     & $\xi\in[\bar\xi,1]$ & $\LevB$ & $11\times11$, degree $5$\\
    \end{tabular}
  \end{ruledtabular}
  \caption{The ten vertex inequalities of $\Poly_\theta$ and their uniform certificates. The last column gives the size of $W'$ and the degree of its entries in $\xi$. Each row describes the unprimed vertex. Its mirror image, when different, has the mirrored saturated strategies and uses the same certificate, transformed by the mirror symmetry.}
  \label{tab:fam-sos}
\end{table}

\begin{table}[t]
  \begin{ruledtabular}
    \begin{tabular}{lccccc}
      NPA Level                               & $k=1$               & $k=3$               & $k=5$               & $k=7$               & $k=8$\\
      \hline
      $1{+}A{+}B$                         & $0.36$              & $0.26$              & $0.18$              & $0.12$              & $0.086$\\
      $1{+}A{+}B{+}AB$                    & $7.0{\cdot}10^{-3}$ & $1.2{\cdot}10^{-2}$ & $1.5{\cdot}10^{-2}$ & $1.0{\cdot}10^{-2}$ & $9.1{\cdot}10^{-3}$\\
      $1{+}A{+}B{+}AA'{+}BB'{+}AB$        & $6.8{\cdot}10^{-3}$ & $9.9{\cdot}10^{-3}$ & $9.9{\cdot}10^{-3}$ & $1.7{\cdot}10^{-3}$ & $3.3{\cdot}10^{-4}$\\
      $1{+}A{+}B{+}AB{+}ABB'$             & $3.4{\cdot}10^{-4}$ & $6.8{\cdot}10^{-4}$ & $2.1{\cdot}10^{-3}$ & $1.5{\cdot}10^{-3}$ & $8.9{\cdot}10^{-4}$\\
      $\LevA$                             & 0                   & 0                   & 0                   & 0                   & 0\\
      $\LevB$                             & 0                   & 0                   & 0                   & 0                   & 0\\
      $1{+}A{+}B{+}AA'{+}BB'{+}AB{+}ABB'$ & 0                   & 0                   & 0                   & 0                   & 0\\
    \end{tabular}
  \end{ruledtabular}
  \caption{Largest value of $\omega_q^{\rm NPA}-1$ over the vertices of $\Poly_\theta$, for $\theta=\pi/8+k\pi/64$ (MOSEK~\cite{mosek} via ncpol2sdpa~\cite{wittek2015}). The entry ``0'' means $|\omega_q^{\rm NPA}-1|<2\times10^{-8}$, the solver accuracy.}
  \label{tab:fam-npa}
\end{table}


\subsection{Identification of the orthogonal faces $\vP_\theta^\perp$}

\begin{theorem}[Orthogonal faces of $\Pth$]\label{thm:fam}
  Let $\theta\in(\pi/8,\pi/4]$. Define $\bar\theta\coloneqq\tfrac12\arcsin\bar s\simeq0.6445$, where $\bar s$ is the root of $2s^3+1=\sqrt2\,(1+s)$ in $(1/\sqrt2,1)$.
  \begin{enumerate}
    \item[(i)] The orthogonal face of $\Pth$ is
      \begin{equation}
        \Pth^\perp = \big\{ \bbeta_\theta(r_0,r_1)
 :\ (r_0,r_1) \in \Poly_\theta \big\}.
      \end{equation}
    \item[(ii)] $\Poly_\theta$ is a hexagon for $\theta\le\bar\theta$ and an octagon for $\theta>\bar\theta$. Its vertices are intersections of the lines of two deterministic strategies.
  \end{enumerate}
\end{theorem}

\begin{proof}
By \cref{thm:fam-sos}, $S_{\bbeta_\theta(V)}\preceq\openone$ for every vertex $V$ of $\Poly_\theta$, so $\omega_q(\bbeta_\theta(V))\le1$. Since $\bbeta_\theta(V)\cdot\Pth=1$, we get $\omega_q(\bbeta_\theta(V))=1$ and $\bbeta_\theta(V)\in\Pth^\perp$. By convexity, the full polytope of inequalities with parameters in $\mathcal{P}_\theta$ is in $\Pth^\perp$. Conversely, by Proposition~\ref{prop:fam-plane}, every $\bbeta\in\Pth^\perp$ is of the form $\bbeta_\theta(r_0,r_1)$, and $(r_0,r_1)\in\Poly_\theta$ since $\Q^*\subseteq\L^*$. This proves (i), and (ii) is a direct consequence of the characterization of $\mathcal{P}_\theta$ obtained in Proposition~\ref{prop:fam-polygon}.
\end{proof}

\subsection{NPA relaxations and numerical methods}\label{app:fam-npa}

For an arbitrary nonlocal extremal point of $\mathcal{Q}$, computing the orthogonal face inside the (at most) two-dimensional plane identified by small variation of the realization parameters (see Appendix~\ref{app:dimensionbound}) can be approached from the inside using numerical implementation of the NPA hierarchy~\cite{NPA07,NPA08} at finite levels. 

We discuss here how this numerical approach compares to the analytical result. Table~\ref{tab:fam-npa} lists the largest excess of the bound over one among the extremal inequalities of the orthogonal face, corresponding to the vertices of $\Poly_\theta$.

\begin{figure}[b]
  \centering
  \includegraphics[width=0.9\textwidth]{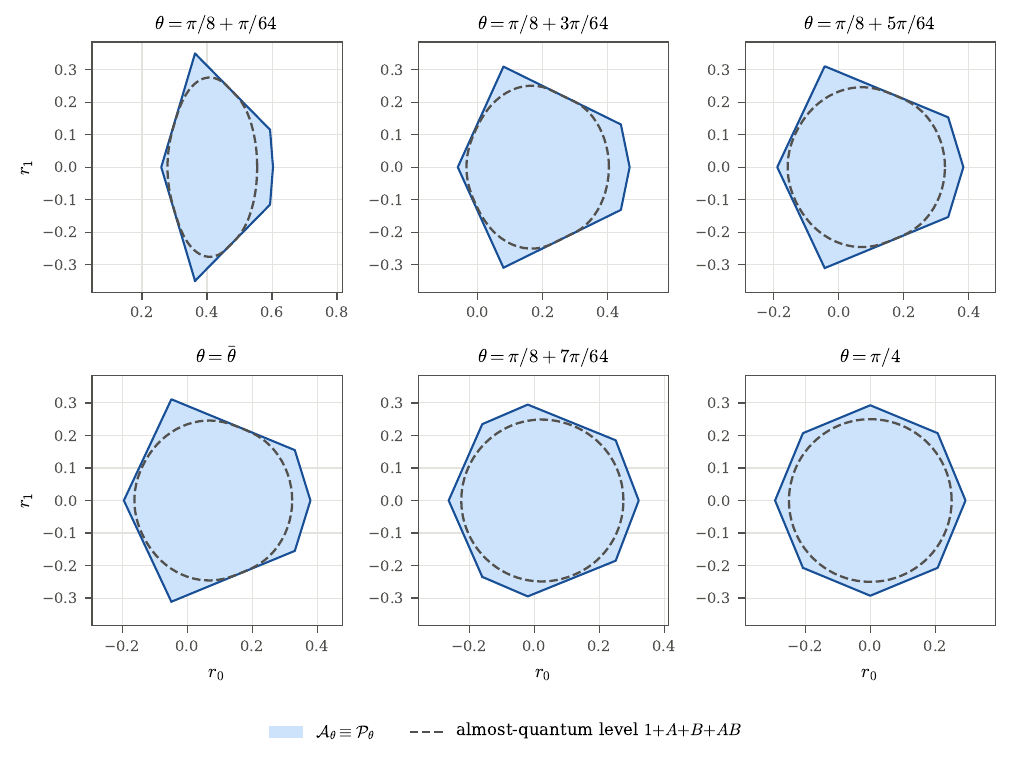}
  \caption{Region certified by the almost-quantum level $\mathcal T_{1+A+B+AB}$ (dashed), compared with the exact section $\Poly_\theta$ (shaded), for six values of $\theta$ ranging  from $\pi/8+\pi/64$ to $\pi/4$. At $\theta=\pi/4$, the dashed region is the disk of radius $1/4$. The region boundaries were found by bisection along 180 rays from the centroid of $\Poly_\theta$, with tolerance $10^{-7}$ on the bound.}
  \label{fig:fam-npa}
\end{figure}

For the tested values $\theta<\pi/4$ the first level certifies quantum bound $1$ for no inequality of the plane. Over the whole plane, its bound stays above one, with minimum $1.24$ at $\theta=0.442$ and $1.13$ at $\theta=0.638$. At $\theta=\pi/4$ it certifies only the CHSH inequality, as already shown in Ref.~\cite{Barizien24-extremal}. The almost-quantum level $\mathcal T_{1+A+B+AB}$~\cite{navascues2015} certifies a convex region strictly inside $\Poly_\theta$. Fig.~\ref{fig:fam-npa} compares the region certified at this level with $\Poly_\theta$ for six values of $\theta$. The level $\mathcal T_{1+A+B+AB+ABB'}$ reaches some vertices but not all. At $\theta=\pi/4$ it reaches the four vertices on the axes and misses the four diagonal ones by $8.9\times10^{-4}$. These need the products $AA'B$ obtained by exchanging the parties. NPA level 3 reproduces $\Poly_\theta$ for all tested $\theta$, showing consistency with Theorem~\ref{thm:fam}. NPA bounds were computed with ncpol2sdpa~\cite{wittek2015} and MOSEK~\cite{mosek}. 

\newpage

\section{Proof of \Cref{thm:not-self-dual}}
\label{app:not-self-dual}

\begin{lemma}
\label{lem:V2-beta}
    Consider the extremal correlation $\vP_{\theta}$ defined in~\cref{eq:fam-P} with $\theta = \bar\theta$ (see \Cref{thm:fam}).
    The Tsirelson inequality $\vb^\star \in \vP_{\bar\theta}^\perp$ that corresponds to the vertex $V_2$ defined in~\cref{prop:fam-polygon} is extremal in $\Q^*$ and its orthogonal face has dimension at least $3$, i.e.~$\dim(F_{\vb^\star})\geq 3$.
\end{lemma}
\noindent Explicitly, let $\bar s\coloneqq\sin2\bar\theta\simeq0.9606$ be the root of $2s^3+1=\sqrt2\,(1+s)$ in $(1/\sqrt2,1)$, and $\bar c\coloneqq\cos2\bar\theta=\sqrt{1-\bar s^2}$. With
\begin{equation}
  \tau \coloneqq 1+2\bar s-2\sqrt2\,\bar s^2, \qquad
  \mu \coloneqq \frac{3-\tau}{2}-\sqrt2+\Big(1-\frac{1}{\sqrt2}\Big)\bar s, \qquad
  \nu \coloneqq \bar s^2-\Big(1-\frac{1}{\sqrt2}\Big)(1+\bar s),
\end{equation}
the inequality reads
\begin{equation}\label{eq:beta-star}
  \begin{aligned}
    \vb^\star\cdot\vP ={}& -\frac{\bar c}{2}\big(1-(1+\sqrt2)\tau\big)\big(\braket{A_0}+\braket{A_1}\big)
      - \mu\big(\braket{A_0}-\braket{A_1}\big)
      + \bar c\,(1-\tau)\braket{B_0} + \tau\braket{B_1}\\
    &+ \frac{1-\tau}{2}\big(\braket{A_0B_0}+\braket{A_1B_0}\big)
      - \frac{\bar c\,\tau}{\sqrt2}\big(\braket{A_0B_1}+\braket{A_1B_1}\big)
      + \nu\big(\braket{A_0B_1}-\braket{A_1B_1}\big) \;\le\; 1 .
  \end{aligned}
\end{equation}
Numerically, $\tau\simeq0.3113$, $\mu\simeq0.2115$ and $\nu\simeq0.3485$, so that
\begin{equation}
  \begin{aligned}
    \vb^\star\cdot\vP &\simeq -0.2460\,\braket{A_0} + 0.1769\,\braket{A_1} + 0.1915\,\braket{B_0} + 0.3113\,\braket{B_1}\\
    &\quad + 0.3443\,\braket{A_0B_0} + 0.2873\,\braket{A_0B_1} + 0.3443\,\braket{A_1B_0} - 0.4097\,\braket{A_1B_1} .
  \end{aligned}
\end{equation}
Its quantum and local bounds are both equal to $1$. The bound is reached by $\vP_{\bar\theta}$ and by the deterministic correlations $\detI(+,+)$, $\detII(-,+)$ and $\detI(-,+)$, while the other thirteen deterministic correlations give at most $0.78$.
\begin{proof}
    According to~\cref{thm:fam}, $\vb^\star$, which is a vertex of $\vP_\theta^\perp$, is extremal in $\vP_{\bar\theta}^\perp$. 
    We show that it is also extremal in $\Q^*$. 
    Suppose $\vb^\star = \lambda \vb' + (1-\lambda)\vb''$ for some $\vb', \vb'' \in \mathcal{Q}^*$ and $\lambda\in(0,1)$. 
    Since $\vb^\star \cdot \vP_{\bar \theta} = 1$, we must have $\vb' \cdot \vP_{\bar \theta} = \vb'' \cdot \vP_{\bar \theta} = 1$, which implies $\vb', \vb'' \in \vP_{\bar\theta}^\perp$.
    Similarly, since $\vb^\star \cdot \Vec{D} = 1$ for all three deterministic points $D \in \{\detI(+,+)$, $\detII(-,+)$ and $\detI(-,+)\}$, this also holds for $\vb', \vb''$. 
    However, $\vb^\star$ is the only point in $\vP_{\bar\theta}^\perp$ maximized by all three deterministic points, so $\vb'=\vb''=\vb^\star$, proving that $\vb^\star$ is extremal in $\mathcal{Q}^*$.
    The same argument shows that, for every $\theta\in(\pi/8,\pi/4]$, every vertex of $\vP_\theta^\perp$ is extremal in $\Q^*$, since it is the only point of $\vP_\theta^\perp$ maximized by the deterministic points whose lines meet there.
    
    Based on~\cref{prop:fam-polygon}(iii), $\vb^\star$ is maximized by the correlation $\vP_{\bar\theta}$ defined in~\cref{eq:fam-P}, and the deterministic correlations $\detI(+,+)$, $\detII(-,+)$ and $\detI(-,+)$.
    We list each of their marginal terms row by row in the following matrix:
    \begin{equation}
    \begin{pmatrix}
    1 & 1 & 1 & 1 \\
    -1 & 1 & 1 & -1 \\
    -1 & -1 & -1 & 1 \\
    \frac{\cos(2\theta)}{\sqrt{2}} & 
    \frac{\cos(2\theta)}{\sqrt{2}} & 
    \cos(2\theta) & 0
    \end{pmatrix}.
    \end{equation}
    Its determinant is $-2(2-\sqrt{2})\cos(2\theta)$, which is nonzero for $\theta \in (0, \pi/4)$, including $\theta = \bar\theta$.
    Hence, the four correlations are linearly independent, so their convex hull has affine dimension $3$. Since they all belong to $F_{\vb^\star}$, so does their convex hull, showing that $\dim(F_{\vb^\star}) \geq 3$.
\end{proof}

\begin{lemma}
\label{lem:orth-face-LD}
    Deterministic correlations in the minimal scenario are extremal in $\mathcal{Q}$ and have orthogonal dimension $7$. 
\end{lemma}
\begin{proof}
    Deterministic correlations are extremal because all their components are $\pm 1$. Now, consider a local deterministic correlation $\vL$ and 8 Bell functionals $\vb_i \in \mathbb{R}^8$.
    Define $\vb_i$ such that all entries are zero except for the $i$-th entry, being the same as the $i$-th entry of $\vL$, which is either $1$ or $-1$.
    Then, $\vb_i \cdot \vL = 1$ for all $i$s. Hence they all lie in $\vL^\perp$.
    The seven vectors $\vb_i - \vb_1$, for $i = 2, \ldots, 8$, are linearly independent. So $\dim\vL^\perp\geq7$. Since $\vL^\perp$ lies in the hyperplane $\{\vb:\vb\cdot\vL=1\}$, we also have $\dim\vL^\perp\leq7$.
\end{proof}

\begin{lemma}
\label{lem:positivities}
    Positivity constraints $- a\langle A_x\rangle - b\langle B_y\rangle - ab\langle A_xB_y\rangle \leq 1$ for all $a,b\in\{\pm1\}$ and $x,y\in\{0,1\}$ are extremal in $\mathcal{Q}^*$ and have orthogonal dimension $7$. 
\end{lemma}

\begin{proof}
    These Bell functionals are extremal in $\mathcal{L}^*$ and thus in $\mathcal{Q}^*$. Furthermore, each is maximized by 8 linearly independent deterministic strategies, showing that their orthogonal dimension is $7$.
\end{proof}

\begin{proposition}(\citep[Proposition 4]{Barizien24-extremal}).
\label{prop:orthdimpreserving}
Any invertible linear map $T:\mathbb{R}^n \to \mathbb{R}^n$ such that $T(K)=K'$ for two convex bodies $K,K' \subset \mathbb{R}^n$ maps extremal points of $K$ to extremal points of $K'$ with the same orthogonal dimension.
\end{proposition}

\noindent We are now ready to prove~\Cref{thm:not-self-dual}.

\begin{proof}[Proof of \Cref{thm:not-self-dual}]

The 16 positivity constraints (see~\Cref{lem:positivities}) and the inequality $\vb^\star$ (see~\Cref{lem:V2-beta}) are a set of 17 extremal points of $\Q^*$ with orthogonal dimension at least $3$. 
Note that $\vb^\star$ is not a positivity constraint, as it is maximized by $\vP_{\bar \theta}$, which saturates none of the positivity constraints.
On the contrary, only $16$ extremal points of $\Q$ with orthogonal dimension at least $3$ exist (the deterministic strategies, see \Cref{lem:orth-face-LD}) since \Cref{thm:dual} ensures that all other (nonlocal) extremal points of $\Q$ have orthogonal dimension at most $2$.
A linear map $T$ with $T(\Q)=\Q^*$ is invertible, since its image contains $\Q^*$, which has nonempty interior.
\Cref{prop:orthdimpreserving} thus ensures that no linear map $T:\mathbb{R}^8 \to \mathbb{R}^8$ satisfies $T(\Q) = \Q^*$.
\end{proof}

\end{document}